\documentclass[10pt]{article}
\usepackage{amsmath}
\usepackage{graphicx}
\usepackage{amsfonts}
\usepackage{amssymb}
\usepackage{epsf}
\usepackage{latexsym}

\def\be{\begin{equation}}
\def\ee{\end{equation}}
\def\bea{\begin{eqnarray}}
\def\eea{\end{eqnarray}}
\def\fn{\footnote}
\def\q{\underline{q}}
\def\p{\underline{p}}
\def\a{\underline{\mbox{a}}}
\def\b{\underline{\mbox{b}}} 
\def\c{{\mbox{c}}} 
\def\aa{{\mbox{a}}} 
\def\bbb{{\mbox{b}}} 
\def\sa{{\mbox{\scriptsize a}}}
\def\sb{{\mbox{\scriptsize b}}} 
\def\sc{{\mbox{\scriptsize c}}} 
\def\r{\underline{r}}
\def\u{\underline{u}}
\def\uv{\underline{v}}

\def\S{\underline{S}}
\def\R{\underline{R}}
\def\P{\underline{P}}
\def\L{\underline{L}}
\def\st{\mbox{\sffamily T}} 
 
\def\Sac{\mbox{\sffamily S}} 
\def\sv{\mbox{\sffamily V}}
\def\se{\mbox{\sffamily E}} 
\def\sL{\mbox{\sffamily L}} 
\def\W{\mbox{\sffamily W}}
\def\sH{\mbox{\sffamily H}}
\def\sW{\mbox{\sffamily W}}
\def\ssW{\mbox{\sffamily{\scriptsize W}}}
\def\ssA{\mbox{\sffamily{\scriptsize A}}}
\def\si{\mbox{\sffamily I}} 
\def\ba{\bar{a}} 
\def\bb{\bar{b}} 
\def\bc{\bar{c}}

\def\ip{i^{\prime}}
\def\jp{i^{\prime}}
\def\kpp{i^{\prime\prime}}

\def\ml{\mbox{\tiny l}}
\def\Tr{\mbox{\scriptsize T}}
\def\sS{\mbox{\scriptsize S}}
\def\tS{\mbox{\tiny S}}
\def\ss{\mbox{\scriptsize s}}
\def\N{\mbox{N}}
\def\sN{\mbox{\scriptsize N}}
\def\S{\mbox{S}}\def\s{\mbox{s}}

\def\sH{\mbox{\scriptsize H}}
\def\uH{\underline{\mbox{H}}}
\def\usH{\underline{\mbox{\scriptsize H}}}
\def\tH{\mbox{\tiny H}}
\def\sL{\mbox{\scriptsize L}}
\def\uL{\underline{\mbox{L}}}

\def\xH{\mbox{H}}
\def\xL{\mbox{L}}
\def\pa{\partial}
\def\d{\textrm{d}}
\def\cr{\mbox{\scriptsize{\bf $\times$}}} 
\def\crr{\mbox{\scriptsize{\bf $\mbox{ } \times \mbox{ }$}}}

\def\B{\mbox{\tiny B}}

\begin{document}
\begin{titlepage}
\begin{center}

\vspace{.3in}

{\Huge{\bf RELATIONAL PARTICLE MODELS.}}

\vspace{.15in} 

{\large {\bf II. USE AS TOY MODELS FOR QUANTUM GEOMETRODYNAMICS}}

\vspace{.3in}

{\Large{\bf Edward Anderson}}

\vspace{.3in}

\noindent{\em Department of Physics, P-412 Avadh Bhatia Physics Laboratory,} 

\noindent{\em University of Alberta, Edmonton, Canada, T6G2J1;}   

\noindent{\em Peterhouse, Cambridge, U.K., CB21RD;}

\noindent{\em DAMTP, Centre for Mathemetical Sciences, Wilberforce Road, Cambridge, U.K., CB30WA.}

\end{center}

\vspace{.3in}


\begin{abstract}

Relational particle models are employed as toy models 
for the study of the Problem of Time in quantum geometrodynamics.  
These models' analogue of the thin sandwich is resolved.  
It is argued that the relative configuration space and shape space of these models 
are close analogues from various perspectives of superspace and conformal superspace respectively.  
The geometry of these spaces and quantization thereupon is presented.   
A quantity that is frozen in the scale invariant relational 
particle model is demonstrated to be an internal time in a certain portion of the 
relational particle reformulation of Newtonian mechanics.  
The semiclassical approach for these models is studied as an emergent time resolution for 
these models, as are consistent records approaches.  
  
\end{abstract}


\vspace{.3in}





\noindent{\bf PACS numbers 04.60-m, 04.60.Ds}

\end{titlepage}

\section{Introduction}

Attempts at quantizing general relativity (GR) 
significantly underly Wheeler's many-routes perspective 
\cite{WheelerGRT, Battelle, MTW, HKT, Phan}, in which   
Einstein's \cite{Ein} traditional `spacetime' route to GR 
is viewed not as {\sl the} route to GR but as {\sl the first} route to GR,  
a theory which is additionally reachable along a number of a priori unrelated routes.  
E.g. six routes to GR are listed in \cite{MTW}.    
Of these, the second (Einstein--Hilbert action) route,  
and the third and fourth routes (the two-way process between this and the 
below geometrodynamical formulation) are relevant to this article.    
The idea is then that some such routes may facilitate quantization, 
while different routes may lead to inequivalent quantum theories.  
This article contemplates variants of the so-called traditional-variables canonical approach 
to quantum gravity (see e.g. \cite{Kuchar80, Kuchar92, POTlit2, Kiefer}).  

Consider the 3 + 1 split Lagrangian for spatially compact without boundary GR\fn{Lower 
and upper-case Greek letters are used as 3-d space and 4-d spacetime indices respectively.}
\be
\si = \int\d\lambda\int\d^3x\sqrt{\gamma}\alpha(\Lambda + \mbox{\Large $\rho$} 
+ \kappa_{\alpha\beta}\kappa^{\alpha\beta} - \kappa^2) \mbox{ } ,
\ee
where $\gamma_{\alpha\beta}$ is the spatial 3-metric 
with determinant $\gamma$, covariant derivative $\nabla_{\alpha}$ and Ricci scalar {\Large $\rho$}, 
$\Lambda$ is the cosmological constant, 
$\alpha$ is the lapse function (proper time elapsed between neighbouring spatial slices), 
$\beta_{\alpha}$ is the shift function (shift in spatial coordinates between neighbouring spatial 
slices used to make that identification), 
$\lambda$ is label time, 
$\kappa_{\alpha\beta}$ is the extrinsic curvature 
$-\frac{1}{\alpha}(\dot{\gamma}_{\alpha\beta} - \pounds_{\beta}\gamma_{\alpha\beta})$, 
the dot is $\pa/\pa\lambda$ 
and $\pounds_{\beta}$ is the Lie derivative with respect to $\beta_{\alpha}$.  
Denoting the gravitational momenta by $\pi^{\alpha\beta}$, 
the corresponding Hamiltonian is \cite{DiracADM}     
\be
\mbox{\sffamily{H}} = 
\int\d\lambda\int \d^3x\sqrt{\gamma}(\alpha{\cal H} + \beta^{\alpha}{\cal M}_{\alpha}) 
\mbox{ } ,
\ee
\be
{\cal M}_{\alpha} \equiv -2\nabla_{\beta}{\pi^{\beta}}_{\alpha} = 0 
\mbox{ } \mbox{ (momentum constraint) } ,
\label{mom} 
\ee
\be
{\cal H} \equiv \frac{1}{\sqrt{\gamma}}
\left(
\gamma_{\alpha\gamma}\gamma_{\beta\delta} - \frac{1}{2}\gamma_{\alpha\beta}\gamma_{\gamma\delta}
\right)
\pi^{\alpha\beta}\pi^{\gamma\delta} - \sqrt{\gamma}(\mbox{\Large $\rho$} + \Lambda) 
\mbox{ } \mbox{ (Hamiltonian constraint) } .  
\label{ham}
\ee

How should this dynamical formulation of GR be interpreted?  
Wheeler explored many possibilities.     
His original attempt, regarded as failed \cite{Geometrodynamica}, 
was this dynamics in vacuo as a theory of everything.  
In later interpretations, source terms have been considered a necessity, and fortunately are manageable.  
Wheeler's next attempt was the {\it thin sandwich approach}.
Baierlein, Sharp and Wheeler \cite{BSW} regarded $\gamma_{\alpha\beta}$ and 
$\dot{\gamma}_{\alpha\beta}$ as knowns 
(i.e. the `thin' limit of taking the bounding `slices of bread' 
$\gamma_{\alpha\beta}^{(\lambda = 1)}$ and 
$\gamma_{\alpha\beta}^{(\lambda = 2)}$ as knowns) 
and solved for the spacetime `filling' in between, in analogy 
with the QM set-up of transition amplitudes between states at two different times \cite{WheelerGRT}.    
They then eliminated $\alpha$ from its own multiplier equation to obtain  
\be
\si = \int\d\lambda\int\d^3x\sqrt{\gamma}\sqrt{(\Lambda + \mbox{\Large $\rho$})
\st_{\mbox{\scriptsize GR}}      }
\ee
for $\st_{\mbox{\scriptsize GR}} = 
\left(
\gamma^{\alpha\gamma}\gamma^{\beta\delta} - \gamma^{\alpha\beta}\gamma^{\gamma\delta}
\right)
(\dot{\gamma}_{\alpha\beta} - \pounds_{\beta}\gamma_{\alpha\beta})
(\dot{\gamma}_{\alpha\beta} - \pounds_{\beta}\gamma_{\gamma\delta})$. 
Then the momentum constraint is replaced by the $\beta_{\alpha}$-variation equation, 
which is furthermore regarded as an equation to be solved for $\beta_{\alpha}$ itself.   
Unfortunately the p.d.e. involved in this attempted elimination ({\it thin sandwich conjecture}) 
is difficult and not much is known about it.    
Counterexamples to its solubility have been found \cite{counterTS}, 
while good behaviour {\sl in a restricted sense} took many years to establish \cite{BF}.  

The more standard interpretation, however, 
is that (\ref{mom}) signifies the irrelevance 
of the coordinatization of $\gamma_{\alpha\beta}$ 
so that one is dealing with the {\sl other} 3 pieces of information in $\gamma_{\alpha\beta}$, 
namely those which represent the `3-geometry' ${\cal G}^{(3)}$.  
Thus GR can be viewed as {\it geometrodynamics}, 
for which superspace = Riem/Diff = \{space of ${\cal G}^{(3)}$'s\} 
is a more appropriate configuration space than Riem.\fn{
Riem = \{space of $\gamma_{\alpha\beta}$'s on a fixed 3-space manifold \}  
and Diff are the 3-space diffeomorphisms.}  
DeWitt and Fischer 
\cite{DeWitt67, DeWitt70, Fischer70} determined that 
superspace is a curved {\it stratified} manifold that has singular boundaries.  
There are various suggestions as to what happens when the dynamical path strikes a boundary 
\cite{DeWitt70, Fischer86}.   
Wheeler \cite{Battelle} emphasized that the 
remaining major remaining conceptual hurdle concerns 
the Hamiltonian constraint ${\cal H}$.  
It is this which naturally provides for superspace a 
(pointwise) metric $\Gamma_{\alpha\beta\gamma\delta}$: 
the prefactor of $\pi^{\alpha\beta}\pi^{\gamma\delta}$ in (\ref{ham}).    
Wheeler furthermore asked why ${\cal H}$ takes the form it does 
and whether this could follow from first principles (`seventh route' to GR) 
rather than from mere rearrangement of the Einstein equations.  
To date, there are two different `seventh routes' which
tighten wide classes of ans\"{a}tze down to the GR form. 
These are from deformation algebra first principles \cite{HKT, Teitelboim}, 
and the `relativity without relativity' approach \cite{RWR, AB, Van, Than, Lan, Phan} 
whereby relational (Machian) first principles lead to the {\sl recovery} of the Baierlein--Sharp--Wheeler  
action of GR.   

What of quantum geometrodynamics?  
One strategy is to attempt superspace quantization in the configuration representation 
$\hat{\gamma}_{\alpha\beta} = \gamma_{\alpha\beta}$ and $\hat{\pi}^{\alpha\beta} = 
-i\hbar\delta/\delta \gamma_{\alpha\beta}$, 
ordering the quantum constraints as follows:
\be
\hat{{\cal M}}_{\alpha}\Phi = \nabla_{\alpha}\frac{\delta}{\delta \gamma_{\alpha\beta}}\Phi = 0
\mbox{ } ,
\label{MOM}
\ee
\be
\hat{{\cal H}}\Phi = 
\left(
\Gamma_{\alpha\beta\gamma\delta}
\frac{\delta}{\delta \gamma_{\alpha\beta}}
\frac{\delta}{\delta \gamma_{\alpha\beta}} - \sqrt{\gamma}(\mbox{\Large $\rho$} + \Lambda)
\right)\Phi = 0 \mbox{ } \mbox{ }
\mbox{ (Wheeler--DeWitt equation) } .  
\label{HAM}
\ee
(\ref{MOM}) is interpreted to mean that the {\it wavefunction of the universe},  
$\Phi$, is a functional on superspace alone: $\Phi[{\cal G}^3]$.  
(\ref{HAM}) is rather more controversial.  
Prima facie, it appears to say that the universe as a whole is `frozen in time' 
or bereft of dynamics (see e.g. \cite{B94I, B94II}), since it looks like 
a time-independent Schr\"{o}dinger equation (TISE) $\hat{H}\Phi = 0$ and not 
a time-dependent one (TDSE) $\hat{H}\Phi = i\hbar\pa\Phi/\pa \tau_?$ 
for $\tau_?$ some notion or other of time.  
However, (\ref{HAM}) is also not well-defined/regularized.  
Need such a poorly-understood equation in truth be a TISE rather than a TDSE?  
 
One possible answer is that there is actually an {\it internal time} hidden within 
the theory.   
This is based on the hope \cite{Kucharearlyit, Kuchar92, POTlit2} 
that a canonical transformation 
\be
\left( 
\gamma_{\alpha\beta}(x), \pi^{\alpha\beta}(x)
\right) 
\longrightarrow 
\left(
\mbox{\Large $\gamma$}_{\Delta}(x), \Pi^{\Delta}(x); 
\gamma^{\mbox{\scriptsize true\normalsize}}_{\mbox{\sffamily\scriptsize Z\normalfont\normalsize}}(x), 
\pi^{\mbox{\scriptsize true\normalsize}\mbox{\sffamily\scriptsize Z\normalfont\normalsize}}(x) 
\right)
\ee
may be found which separates out 4 embedding variables $\mbox{\Large $\gamma$}_{\Delta}$ 
from the 2 true degrees of freedom of GR  
$\gamma^{\mbox{\scriptsize true\normalsize}}_{\mbox{\sffamily{\scriptsize Z\normalfont\normalsize}}}(x)$.  
Then in these new variables, the constraints have been molded into the form 
\be
\Pi_{\Delta}(x) = \Pi_{\Delta} 
\left[ 
\gamma^{\mbox{\scriptsize true\normalsize}}_{\mbox{\sffamily\scriptsize Z\normalfont\normalsize}}(x)
, \pi^{\mbox{\scriptsize true\normalsize}}_{\mbox{\sffamily\scriptsize Z\normalfont\normalsize}}(x) ; 
\mbox{\Large $\gamma$}_{\Delta}(x)
\right] 
\equiv  
- {\cal H}_{\Delta}^{\mbox{\scriptsize true\normalsize}}
\left[
\gamma^{\mbox{\scriptsize true\normalsize}}_{\mbox{\sffamily\scriptsize Z\normalfont\normalsize}}(x)
, \pi^{\mbox{\scriptsize true\normalsize}}_{\mbox{\sffamily\scriptsize Z\normalfont\normalsize}}(x) ; 
\mbox{\Large $\gamma$}_{\Delta}(x), \Pi^{\Delta}(x)
\right] 
\mbox{ } , 
\ee
after which rearrangement and use of the configuration representation one clearly obtains 
a TDSE at the quantum level after all (the linear momentum term providing the time derivative).  

To provide a salient example of internal time candidate, 
I first consider a further classical rearrangement.
Superspace still has 3 degrees of freedom per space point 
while GR has two due to ${\cal H}$.  
A question then is whether there is a `natural 2/3 of superspace'.
There is: conformal superspace =
$\mbox{Riem}/(\mbox{Diff} \times \mbox{Conf}) = 
\{\mbox{Space of } {\cal C}^{(2)}\mbox{'s}: \mbox{ conformal 3-geometries}\}$.\fn{Conf are 
the conformal transformations.  See \cite{CSlit, FischMon} for studies of conformal superspace.}    
This second arena furbishes a set of `eighth routes': the conformogeometrodynamical alternatives 
of York \cite{Yorklit, YT, POTlit2} 
(see \cite{ABFKO} for a recent relational derivation), 
later explicitly favoured by Wheeler \cite{Wheelerlater} over the above approaches, 
and the more recent \cite{CTS} (conformal thin sandwich). 
These approaches originate from Lichnerowicz's study \cite{Lich} of maximal slices
\be
\pi \equiv \pi^{\alpha\beta}\gamma_{\alpha\beta} = 0 
\mbox{ } ,  
\label{Max}
\ee 
which York significantly amplified by generalization to constant mean curvature slices
\be
\pi = \pi^{\alpha\beta}\gamma_{\alpha\beta} = {C}{\sqrt{\gamma}}
\mbox{ } .  
\label{CMC}
\ee 
In these approaches, the 4 constraints are {\sl tractable}  
(at least as regards existence, uniqueness and robustness to inclusion of matter \cite{Lichconcat}), 
the hardest part of which is solving 
the conformally-transformed ${\cal H}$ 
\be
8\nabla^2\psi = \left(\frac{\pi^2}{6} + \Lambda \right)\psi^5 + \mbox{\Large $\rho$}\psi - 
\pi_{\alpha\beta}\pi^{\alpha\beta}\psi^{-7}  
\label{LYE}
\ee 
(a Lichnerowicz--York type equation) for the scale factor $\psi$.  

Now, ignoring the solution of the momentum constraint for simplicity of presentation,  
a canonical transformation permits the York time \cite{YT, POTlit2} 
$\tau_{\mbox{\scriptsize Y\normalsize}} 
= \frac{2}{3\sqrt{\gamma}}(\gamma_{\alpha\beta}\pi^{\alpha\beta})$ 
to serve as a coordinate while its conjugate quantity $\sqrt{\gamma}$ is now a momentum.  
Then the Hamiltonian constraint is replaced by 
$\sqrt{\gamma} =$ [the $\psi$ which solves eq. (\ref{LYE})$]^6$.  
As this is linear in the momentum $\sqrt{\gamma}$, this equation becomes a TDSE 
\be
i\frac{\delta\Phi}{\delta \tau_{\mbox{\scriptsize Y\normalsize}}   } = 
- \widehat{\sqrt{\gamma}}\Phi 
\mbox{ } .  
\ee
The obstruction to this particular resolution is that how to solve 
the complicated quasilinear elliptic equation (\ref{LYE}) is not in practice generally known, 
so the functional dependence of ${\cal H}^{\mbox{\scriptsize true}}$ on the other variables is not known, so the  
quantum `true Hamiltonian' $\widehat{{\cal H}}^{\mbox{\scriptsize true}}$ cannot be explicitly defined.
No other internal time candidate is currently known to work in practice either.    

Another possible answer is that perhaps closed universes 
do not have a fundamental notion of time.    
A major issue then is explaining why we appear to experience dynamics.    
Undeniably it is dynamics {\sl of subsystems} that we experience.  
Along the lines of the semiclassical approach \cite{DeWitt67, SCA, SCB, Banks, Zeh, SCAII, Kiefer}, 
the TISE of a system can lead to the emergence of (WKB)-TDSE's for subsystems 
if a number of approximations hold, including subdivision into subsystems of 
`light variables' $\xL_{\mbox{\scriptsize{\sffamily A}}}$ and 
`heavy variables' $\xH_{\mbox{\scriptsize{\sffamily B}}}$ 
that are not entirely negligibly intercoupled and such that the latter has a WKB regime.    
This requires 
\be
\Phi = e^{iM_{\tH}\ssW({\mbox{\scriptsize H}}_{\mbox{\tiny{\sffamily B}}})/\hbar}
\phi(\xH_{\mbox{\scriptsize{\sffamily B}}}, \xL_{\mbox{\scriptsize{\sffamily A}}})
\mbox{ } \mbox{ } \mbox{ (WKB ansatz for the wavefunction) } , 
\ee
where $\W$ is the principal function.  
Then one peels off the Hamilton--Jacobi equation as the leading order terms, 
and moreover keeps the derivative cross term to the next order of approximation.  
It is this that supplies the emergent WKB time $\tau_{\mbox{\scriptsize WKB}}$.  
In the case of `heavy gravitational modes' supplying WKB time to 
`light minimally coupled matter (mcm) modes' 
(which contribute additive portions ${\cal H}^{\mbox{\scriptsize mcm}}$ and 
${\cal H}_i^{\mbox{\scriptsize mcm}}$ to ${\cal H}$ and ${\cal H}_i$ respectively), 
one has the emergent 
\be
i\hbar\frac{\delta\phi}{\delta \tau_{\mbox{\scriptsize WKB}}} = 
\left(
\int \d^3x(\alpha{\cal H}^{\mbox{\scriptsize mcm}} + \beta^{\alpha}{\cal M}_{\alpha}^{\mbox{\scriptsize mcm}})
\right)
\phi 
\label{naga}  \mbox{ } \mbox{ } \mbox{ (Tomonaga--Schwinger WKB-TDSE) } .
\ee

However, in the context of whole universes, 
it is not clear whether these approximations used are appropriate.   
An alternative timeless type of approach being developed \cite{Page, B94II, EOT, Hartle, H99} 
is that in which dynamics/history are to be recovered  
from {\sl consistent records} (particular sorts of instantaneous configurations).      
There is then an additional issue of what in nature causes these to be selected 
rather than instants from which a semblance of dynamics/history cannot be reconstructed.  
Barbour speculates \cite{Bararr, B94II, EOT} that the asymmetry of the 
underlying curved stratified configuration space 
causes concentration of the wavefunction in `time capsules' (his notion of consistent records).  
On the other hand, Gell-Mann--Hartle \cite{GMH} and Halliwell \cite{H99} have conjectured recovery of apparent dynamics for 
subsystems as following from a light--heavy split and decoherence 
(bearing some similarities to techniques of the semiclassical approach).  

%
The difficult and unresolved nature of quantum GR has led to the study of 
a wide range of {\sl toy models} (see e.g. \cite{Kuchar92, Kuchar93, POTlit3, Kiefer}). 
These vary in how they resemble GR, and how closely (such toys ignore some of 
of the following crucial features of GR: 
the infinite dimensionality of the configuration space, 
the linear momentum constraint, 
the high nonlinearity, 
the indefinite-signature kinetic term and 
the presence of a complicated potential term).  
Midisuperspace models \cite{midi} are closer to full GR than the more commonly studied 
minisuperspace \cite{DeWitt67, mini, minimetric, Misref, MTW, KucharRyan} models, 
of course at the price of being harder to study, 
while there are many other sorts of models: 2+1 GR \cite{Carlip},  
small perturbations about minisuperspace \cite{HH}, 
parametrized \cite{toystory, Kuchar92} 
and relational \cite{B94I, B94II, Kuchar92, EOT, Kiefer} particle models, 
and coupled oscillators \cite{Ashtekar, Kuchar92, H99}. 

This paper focuses on two relational particle mechanics (RPM) models as quantum gravity toys: 
the {\it Barbour--Bertotti 1982 (BB82) } RPM \cite{BB82} (see paper {\bf I} 
\cite{I}),\fn{This toy has been used by Smolin \cite{Smolin}, Rovelli \cite{Rovelli}, 
and Barbour and collaborators \cite{BS, B94I, B94II, EOT, RWR}, 
and has been acknowledged as useful in the reviews \cite{Kuchar92, Kiefer}.  
Simpler versions of these toys were used by DeWitt \cite{DeWitt70}, 
and by Brown and York \cite{YB1, YB2}.}  and 
{\it Barbour's scale-invariant particle theory (SIPT)} \cite{SIPP} (see Sec 2),\fn{This 
toy is recent; it has been used in \cite{ABFKO}.}  which has one further constraint\fn{For  
particle number $N$ and dimension $d$, I use 
upper-case Latin indices as particle labels running from 1 to $N$, 
lower-case Latin indices as relational labels from 1 to $N$ -- 1 and 
barred lower-case Latin indices as shape labels running from 1 to $(N - 1)d$ -- 1.}   
\be
{\cal E} \equiv \sum_{I = 1}^{N}\q_I\cdot\p^I = 0 \mbox{ } \mbox{ } \mbox{ (Euler constraint) } 
\label{Euler} 
\mbox{ } .
\ee
My aim is to demonstrate that BB82 RPM and SIPT 
are a fruitful multi-purpose arena for conceptualizing and modelling 
toward understanding geometrodynamical formulations of GR.  
I explain the immediate analogies of the original formulations of BB82 RPM and SIPT  
to the geometrodynamical and maximally sliced conformogeometrodynamical formulations of GR in Sec 2.

Furthermore, these models' {\sl thin sandwich analogues are}  
not conjectures but rather {\sl explicitly soluble}. 
For BB82 RPM this is precisely 
the content of {\bf I}.3, 
while its SIPT counterpart is presented in Sec 3.    
Next (Sec 4), the relative configuration space (RCS) $\Re^{dN}/$Eucl($d$) of BB82 RPM is a 
{\sl simple analogue of superspace}, while shape space $= \Re^{dN}/$Sim($d$) 
is a {\sl simple analogue of conformal superspace}.\fn{Eucl($d$) is the Euclidean group while Sim($d$) 
is the similarity group Eucl($d$) $\times$ Sca for Sca the scale transformations.}
This comes into the study of both SIPT  
(as the configuration space) and  BB82 RPM (as an important quotient space of the RCS, 
and now parallelling the role of conformal superspace in constant mean curvature-sliced GR). 
   
\noindent
These analogies are close enough for fruitful technical work to be done, 
insofar as RCS and shape space are in general curved, stratified manifolds 
capable of having singular boundaries \cite{Gergely, GergelyMcKain} (and thus are  
more sophisticated versions of the toys DeWitt used \cite{DeWitt70}).  
One approach for these models, opened up by their thin sandwich being solvable, 
is to work directly on the reduced configuration spaces.  

In Sec 5 I explain that the method of {\bf I}.6--8 covers 
both full- and semi-RCS analogues of superspace quantization, 
and I develop the shape space analogue of conformal superspace quantization.
Sec 6 I dedicate to the quest for internal time.  
While SIPT is frozen in the Euler quantity ${\cal E}$, 
I demonstrate that {\sl in a certain portion 
of the portion of Newtonian mechanics (NM) covered by the BB82 reformulation, 
${\cal E}$ is monotonic and can serve as an internal time 
in direct analogy to the York time in GR}.  
My finding of this internal time already within this portion of NM 
stripped of its absolute structures is an ironic contribution 
to the classical absolute or relative motion debate, 
and provides cause to be skeptical about aesthetic or philosophical arguments 
as to why it would be unnatural for relational theories such as GR to actually 
have a privileged time.  

Having already worked around Barbour and Smolin's \cite{BS} objections to 
the semiclassical study of RPM's in {\bf I}, 
I study this further in Sec 7.  
{\sl If} the WKB ansatz (associated with classicality of the `heavy subsystem') is adopted,  
apparent dynamics for subsystems within universes governed by a TISE follows.    
But this ansatz currently lacks justification at the level of whole universes (Sec 8.1--3),  
so I also consider more radical consistent records approaches of e.g. Barbour \cite{B94II, EOT} 
and Halliwell \cite{H99} for simple RPM's in Sec 8.4--6.  
I conclude in Sec 9.  
Some of this paper's results were outlined in my Paris talk \cite{Panderson}.

\section{Relational particle models}

Consider the following 3 toy model schemes (Fig 1 A to C) for GR (Fig 1 D).  
\begin{figure}[h]
\centerline{\def\epsfsize#1#2{0.6#1}\epsffile{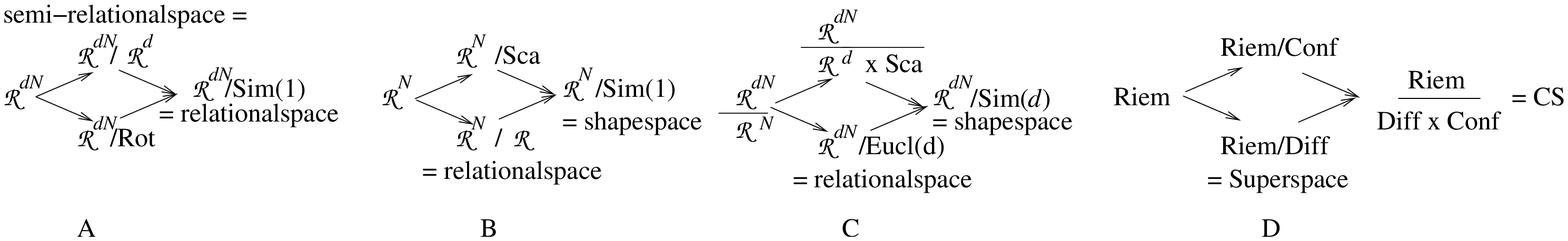}}
\caption[]{\label{TO7.ps}
\footnotesize 
\normalsize}
\end{figure}
These should be judged both by how mathematically accessible they are and by their resemblance to GR.  
Difficulties arise from $d > 1$, quotienting out rotations, and curved configuration spaces 
with edges and special points, while 
$d = 1$, quotienting out translations and flat configuration spaces are simpler (c.f {\bf I}).  
By this token, the principal scheme B of this paper,  
is simpler than the traditional $d = 3$ relational program A, 
while, nevertheless also having curved configuration spaces with edges and special points (see Sec 4).  
Moreover, I argue that scheme C is closest\fn{For $d > 1$ with the translations already taken care of 
at the outset by adopting relative Jacobi coordinates.} to the GR scheme D, 
with the rotations playing the role of the 3-diffeomorphisms 
and the scalings playing the role of the conformal transformations.  
This further motivates study of scheme B as an isolation of some of the interesting features 
of scheme C.  Indeed scheme B viewed in parallel with scheme C in Jacobi coordinates, 
bears close resemblances to the relation between minisuperspace and full GR.  

There are two realizations of B, C.  
The first involves Sim($d$) actually being the redundant motions via one further constraint ${\cal E}$.   
This realization is SIPT.      
The second involves Eucl($d$) being the redundant motions i.e. the traditional RPM  
program for which BB82 provides a concrete realization, 
but now with the variables furthermore being split into one scale and $d(N - 1) - d(d-1)/2 -1$ 
{\sl shape variables}.  
These two realizations furthermore turn out to be heavily inter-related 
both conceptually and technically.

\subsection{BB82-type relational formulation of NM: analogy with GR}

The BB82-type relational formulation (see Sec {\bf I}.2) as a route to NM is analogous to 
the Baierlein--Sharp--Wheeler formulation as a route to GR.\fn{What of other routes to NM?  
Newton's own route is sui generis rather than pairable with any GR route.  
Cartan's bears close parallel to Einstein's traditional route to GR.    
The variational work of Euler and Lagrange bears parallels with both the second and third 
routes to GR, due to the difference between space and time inherent in NM.  
There appears to be no need for, or possibility of, NM counterparts for the fourth to sixth 
routes to GR, as these stem from rearranging {\sl field} equations 
or other {\sl field-theoretic} notions.  
I treat NM counterparts of the `seventh' and `eighth' routes in Sec 2.3--4.} 
The Baierlein--Sharp--Wheeler elimination is the same `reverse route' as 
obtaining the Jacobi principle from the Euler--Lagrange principle.  
In both cases, one gets a quadratic constraint as a primary constraint 
due to the peculiar reparametrization invariant square root form of the action: 
energy conservation for the BB82 formulation and the Hamiltonian constraint for GR.    
In both cases, linear constraints arise from varying with respect to auxiliaries: 
$\mbox{a}_{\alpha}$, $\mbox{b}_{\alpha}$ variation of the BB82 action  
give zero total momentum and angular momentum constraints, 
while $\beta_{\alpha}$ variation of the Baierlein--Sharp--Wheeler action gives the momentum constraint.\fn{The 
analogy can be made even more complete by dressing $\beta_{\alpha}$ as a $\dot{s}_{\alpha}$ 
and $\alpha$ as an emergent $\dot{O}$.}  
The total energy $\se$ is in direct correspondence with the cosmological constant $\Lambda$.  
The differences from GR are in the other features of GR listed on p3, including that the natural potentials for 
RPM's to have are often simpler (and in any case simpler) than those one encounters in GR.

\subsection{Original formulation of SIPT and its analogy with GR}

Next I introduce SIPT, working with all the auxiliary variables and corresponding constraints kept.  
I demand good objects under translations and rotations {\sl and} scalings.  
I view this as choosing a simple alternative spatial geometry to Euclidean:   
similarity geometry.\fn{Geometrodynamics, moreover, can be thought of as arising from the relational 
approach by the alternative choice of a complicated set of spatial geometries.}
I proceed indirectly, using a scaling auxiliary $\c$ alongside the translation and rotation 
auxiliaries $\mbox{a}_{\alpha}$, $\mbox{b}_{\alpha}$.  
The $A$th particle's position $q_{A \alpha}$ is replaced by an arbitrary-frame 
position 
\be
q_{\alpha A}^{\prime} = 
\c\left(q_{A\alpha} - \aa_{\alpha} - {\epsilon_{\alpha}}^{\beta\gamma}\bbb_{\beta}q_{A \gamma}\right)  
\mbox{ } .  
\ee
The nontrivial part of this construction, paralleling the situation in {\bf I}.2, is the construction of the kinetic term.  
Namely, along the lines in \cite{Stewart}, $\pa q_{A\alpha}/\pa\lambda$ is not in itself a tensorial operation under 
$\lambda$-dependent Sim(d) transformations.  
It should rather be seen as the Lie derivative 
$\pounds_{\pa q_{A\alpha}/\pa\lambda}$ in a particular frame, which transforms to the Lie derivative with respect to 
`$\pa q_{A\alpha}/\pa\lambda$ corrected additively by generators of the translations, rotations and scalings', which 
gives the combination 
\be
\&_{\dot{\sa}, \dot{\sb}, \dot{\sc}}q_{A \alpha} = 
\c\left(
\dot{q}_{A \alpha} -\dot{\aa}_{\alpha} - {\epsilon_{\alpha}}^{\beta\gamma}\dot{\bbb}_{\beta}q_{A \gamma} 
+ \frac{\dot{\c}}{\c}q_{A \alpha} 
\right) 
\mbox{ } .
\ee

Then, using that the upstairs spatial indexed $\delta$-tensor itself scales as $c^{-2}$, the kinetic term is 
\be
\st(\q_I, \dot{\q}_J, \dot{\a}, \dot{\b}, \c, \dot{\c} ) = 
\frac{1}{2}\sum_{I = 1}^{N}\delta^{AB}c^{-2}\delta^{\alpha\beta}m_A
(\&_{\dot{\sa}, \dot{\sb}, \dot{\sc}}q)_{A \alpha}
(\&_{\dot{\sa}, \dot{\sb}, \dot{\sc}}q)_{B \beta} \mbox{ } 
\ee
which therefore scales overall as the zeroth power of c.  The Jacobi-type action is 
\be
\Sac(\q_I, \dot{\q}_I, \dot{\a}, \dot{\b}, \mbox{c}, \dot{\mbox{c}}) = 
2\int \d\lambda\sqrt{(\se - \sv(\q_I))\st(\q_I, \dot{\q}_I, \dot{\a}, \dot{\b}, \c, \dot{\c}) } 
\label{SIPTJac} 
\mbox{ } .
\ee

The momenta are now 
\be
p^{\alpha A} = \sqrt{\frac{\se - \sv}{\st}}m_A\delta^{\alpha\beta}
\left(\dot{q}_{\beta A} - \dot{\aa}_{\beta A}  
- {\epsilon_{\beta}}^{\gamma\delta}\dot{\bbb}_{\gamma}q_{\delta A} + \frac{\dot{\c}}{\c}q_{\beta A}
\right) 
\mbox{ } .  
\ee
The square root form of the action continues to give the same primary constraint 
$\overline{{\cal Q}} = {\cal Q} - \se$ ({\bf I}.7), while $\aa_{\alpha}$ and $\bbb_{\alpha}$ variation still gives 
${\cal P}^{\alpha} = 0$ and ${\cal L}^{\alpha} = 0$ 
({\bf I}.8--9); now 
$\c$-variation additionally gives rise to the Euler constraint ${\cal E}$ (eq. \ref{Euler}).  
The Poisson brackets already computed ({\bf I}.10) remain the same, 
while the 4 new brackets involving ${\cal E}$ are:   
\be
\{{\cal E}, {\cal E}\} = 0 
\mbox{ } ,
\{{\cal E}, {\cal P}^{\alpha}\} = {\cal P}^{\alpha} 
\mbox{ } , 
\{{\cal E}, {\cal L}^{\alpha}\} = 0 
\mbox{ } , 
\{{\cal E}, \overline{{\cal Q}}\} = 2\overline{{\cal Q}} -  
\left( 
2(\sv - \se) + \sum_{A = 1}^{N}\q_{A}\cdot\frac{\pa (\sv - \se)}{\pa \q_{ A}} 
\right)
\mbox{ } . 
\ee
Two results follow from these brackets. 
1) unless the Euler condition of homogeneity of degree $-2$, 

\noindent $\q_{A}\cdot{\pa (\sv - \se)}/{\pa \q_{A}} = - 2(\sv - \se)$, 
is obeyed, these brackets do not close.\fn{Note that $\se$ plays a different 
role in SIPT than in NM as a consequence of this.}   
Thus mathematical consistency requires this to be the case. 

Consequently, the action (\ref{SIPTJac}) is overall homogeneous of degree zero, 
and is {\sl naturally a function of ratios alone}.      
2) the moment of inertia $J$ is a conserved quantity.  
I note that 1) and 2), in addition to featuring in Barbour's work, 
are fairly well-known in the theoretical celestial mechanics literature
(see e.g. \cite{Diacu, Montgomery, Montgomery2}), and also in  
the QM literature \cite{Conf} where what is conventionally studied 
is rather {\sl conformal symmetry} of $1/r^2$ potentials in distinction to 
the many-particle, constrained set-up of SIPT.  
I furthermore note that Barbour {\sl creatively combines} 
these two well-known elements in a novel way:  he can strike a close analogy with Newtonian 
physics by using powers of the post-variationally constant $J$ in the potential as 
some of the powers which have to add up to --2. 
This produces worlds which are capable of mimicking more general Newtonian worlds.\fn{However, 
while I consider BB82 RPM as plausible and well-motivated reformulations for at least a portion of 
nonrelativistic classical and quantum dynamics in {\bf I}, 
for the moment I treat SIPT solely as a technically interesting toy model 
for quantum gravity (rather than attempting to find evidence for/a 
killer counterexample against it being a plausible competitor to/a 
technically important regime within standard nonrelativistic classical and quantum theory).}
By removing the overall scale from the counting in Sec {\bf I}.2, 
$N$ and $d$ must be such that $d(2N - d - 1)/2 > 2$ for nontriviality, so 
$N \geq 4$ for 1-$d$.

\subsection{Relational particle models' counterparts of `seventh routes' to GR}

Despite the resemblances of Sec 2.1, note that there is not
a close parallel between the BB82 route to NM 
and the {\sl tightening} aspects of the relativity without relativity (RWR) and 
Hojman--Kuchar--Teitelboim (HKT) deformation algebra `seventh routes' to GR.   
For, that tightening arises from the restrictiveness of the emergent or assumed 
constraint interweaving of GR, while the RPM's have far less constraint interweaving. 
For BB82, the variant of RWR that \'{O} Murchadha and I investigated \cite{OMSan} 
fails here (because it relies on linear constraints being integrabilities of ${\cal H}$, 
while here the analogue $\overline{\cal Q}$ is not interwoven with linear constraints).  
There is some interweaving in SIPT but it is not of the right sort to work in this way either.  
In BB82, the standard RWR technique of using consistency under constraint propagation leads to the 
result that if one sets $\sv$ arbitrary, then one is led to $\sv(|\r_{ab}| \mbox{ alone})$; 
in fact, that's how the original BB82 paper works. 
In SIPT, the RWR technique furthermore gives the homogeneity property of $\sv$.  
No more than this can be gleaned.  
Done the HKT way [regarding in each case a particular algebra, now Eucl($d$) and Sim($d$) respectively,  
as fundamental and demanding that constraint ansatze close in precisely that way], 
for the BB82 formulation I find that 
\be
\sum_{\ssA = 0}^{\infty} f(|\r_{IJ}|)_{I_1...I_{\ssA}}p^{I_1} ...p^{I_{\ssA}}
\label{HKT1}
\ee
survives, while in SIPT furthermore  
$
f_{I_1...I_{\ssA}} = l_{I_1...I_{\ssA}}(\r_{IJ}) k(\r_{IJ})
$
for $l_{I_1...I_{\ssA}}$ homogeneous functions of degree {\sffamily A} and $k$ a homogeneous function of degree $- 2$.
These represent considerable generalizations of BB82 RPM and SIPT, corresponding to more 
complicated Jacobi actions as discussed in \cite{Van} and Sec {\bf I}.5.

\subsection{Relational particle models' counterparts of `eighth routes' to GR}  

SIPT shares the abovementioned parallels of the BB82 formulation with GR, 
{\sl and} its additional constraint ${\cal E} = \sum_{I = 1}^{N} \q_{I} \cdot \p^{I} = 0$ 
parallels the maximal slicing condition $\gamma_{\alpha\beta}\pi^{\alpha\beta} = 0$ of GR.  
While SIPT is manifestly a theory of pure shape, 
BB82 RPM may be recast in shape--scale variables, which furthermore 
now parallels constant mean curvature-sliced GR.\fn{See Secs 4, 6 for this recasting, after which 
further observations of this parallel with constant mean curvature-sliced GR emerge.}

\section{The thin-sandwich prescription works for these models}

Consider BB82 RPM as a first example.   
An entirely valid interpretation of Sec {\bf I}.3 is 
that one is carrying out the thin-sandwich prescription:   
starting from a 
$\Sac_{\mbox{\scriptsize Euler--Lagrange}}(\q_I, \dot{\q}_J, \dot{\a}, {\dot \b}) = \int \d t(\st - \sv)$, 
pass to the Jacobi form, then vary with respect to the auxiliaries 
and then eliminate these from their own variational equations.  
Note that this analogue model resolves its thin sandwich conjecture as this last step 
is tractable algebra.   

As a second example, for SIPT, $\sv$, $\st$ are as in the previous section, 
so the auxiliary $\c$ also requires elimination.  
I let $\zeta = \mbox{log}\mbox{\scriptsize\mbox{ }} \c$ so that only $\dot{\zeta}$ features in the action.  
Then the $\zeta$-variation of the $\Sac^{**}(\r_{IJ}, \dot{\r}_{KL}, \dot{\zeta})$ 
obtained from (\ref{SIPTJac}) by elimination of $\dot{\a}$ and $\dot{\b}$ reads 
\be
\dot{\zeta} = - \frac{1}{J} \sum_{I <}\sum_J m_Im_J\delta^{\alpha\beta}r_{\alpha IJ}r_{\beta IJ} 
\mbox{ } .
\ee  
One arrives at this simple form because the $\c$-corrected $\L_{\alpha}$ takes the form 
$\sum_{I <}\sum_{J} \frac{m_Im_J}{M}{\epsilon_{\alpha}}^{\beta\gamma}    
\r_{\beta IJ}\c(\dot{\r}_{\gamma IJ} + \frac{\dot{\sc}}{\sc}r^{\gamma}_{IJ})$ 
and so the correction terms here vanish by symmetry--antisymmetry in $\beta\gamma$.
Then using this to eliminate $\dot{\zeta}$ gives 
$\Sac^{***}(\r_{IJ}, \dot{\r}_{KL}) = 2\int \d\lambda\sqrt{(\se - \sv)\st^{***}(\r_{IJ}, \dot{\r}_{KL}) }$ 
with   

\noindent
\be
\st^{***}(\r_{IJ}, \dot{\r}_{KL}) = 
\frac{    \Sigma_{\dot{\r}^2_{IJ}}\Sigma_{\r^2_{IJ}} - (\Sigma_{\r_{IJ}\dot{\r}_{IJ}})^2   }
     {    2M\Sigma_{\r_{IJ}}    } 
- \frac{1}{2}\stackrel{\B}{L}_{\alpha}(\stackrel{\B}{I}{}^{-1})^{\alpha\beta}\stackrel{\B}{L}_{\beta}
\label{noaborc}
\ee
where $\Sigma_{uv} \equiv \sum_{I <}\sum_J m_Im_J\u \cdot \uv$, and where 
the barycentric inertia tensor $\stackrel{\B}{I}_{\alpha\beta}$ and  angular momentum 
$\stackrel{\B}{L}_{\alpha}$ are given by ({\bf I}.15).  
$\sum_{I = 1}^{N}\sum_{J = 1}^{N}\sum_{K = 1}^{N}\sum_{L = 1}^{N}m_Im_Jm_Km_L|\r_{IJ}\times \r_{KL}|^2$ is 
an alternative form for the first term of (\ref{noaborc}). 

Furthermore, while it has so far been established that the above can 
be written in terms of relative coordinates $\r_{IJ}$, a fortiori by the overall homogeneity 
of degree 0 of $\sv\st$, the Jacobi action is homogeneous 
of degree zero in the $\r_{IJ}$ and is thus a function just of {\sl shape variables} 
such as, for ${I,J} \neq {K,L}$, $|\r_{IJ}|/|\r_{KL}|$  (ratios) or 
$\mbox{arccos}(\r_{IJ} \cdot \r_{KL}/ |\r_{IJ}||\r_{KL}|)$ (angles) and their velocities.  
While not all the $\r_{IJ}$ or shape variables are independent, an independent 
set of them can be chosen in each case (see Sec 4).  

It should be noted that these elimination moves work in close analogy in dimension 2.  
In dimension 1, they work as closely in analogy as possible, given that there is no rotation.  
To complete the Baierlein--Sharp--Wheeler procedure, given the final computed relational kinetic term in each case, 
one can then construct  
$\d t/\d\lambda = \sqrt{\st_{\mbox{\tiny{Relational}}}/(\se - \sv)}$ 
and substitute this into the action.    
One may next consider combining this with the path integral approach of 
Brown and York \cite{YB2} (which was for an unconstrained version of the Jacobi action) 
to investigate the fruitfulness of the sort of transition amplitude approach that Wheeler 
envisaged as arising from resolving thin sandwich schemes.  
Another application of succeeding in this elimination is working directly on 
RCS and shape space in the next section.

\section{Classical study of configuration spaces for both models}

\subsection{Topology}

The simplest models' configuration spaces are displayed in Fig 1.    
\begin{figure}[h]
\centerline{\def\epsfsize#1#2{0.4#1}\epsffile{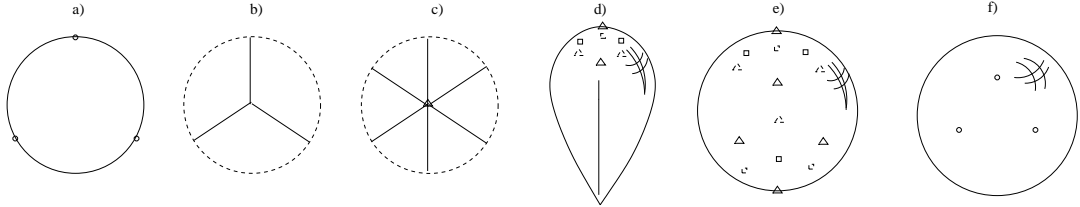}}
\caption[]{\label{TO8.ps}
\footnotesize  
a) The shape space for $N = 3$, $d = 1$ that includes quotienting out reflections.  
The dots are binary collisions.  
b) The RCS for $N = 3$, $d = 1$ that includes quotienting out reflections. 
c) The RCS for $N = 3$, $d = 1$ without quotienting out reflections.  
d) The `$\Re P^2$' shape space for $N = 4$, $d = 1$ that includes quotienting out reflections.  
The triangles denote triple collisions and the squares are simultaneous binary collisions. 
e) The `$S^2$' shape space for $N = 4$, $d = 1$ without quotienting out reflections.  
f) The shape space for $N = 3$, $d = 2$ or $3$.  
Binary collision lines joining various triple and simultaneous binary collisions are omitted from d), e).  
These are displayed in, or deducible from, the construction of these shape spaces described in Fig 3.  
\normalsize}
\end{figure}
%
%
%
\begin{figure}[h]
\centerline{\def\epsfsize#1#2{0.4#1}\epsffile{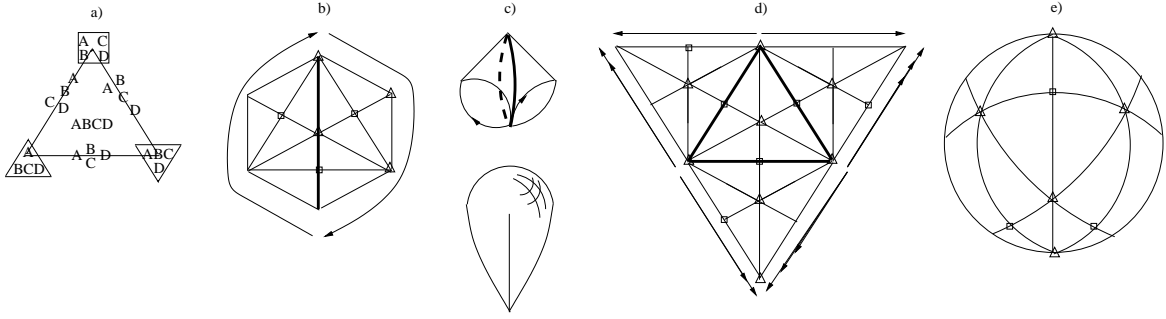}}
\caption[]{\label{TO9.ps}
\footnotesize Constructing representations of shape spaces for 4 particles on a line.  
Let the particles be A, B, C, D, regarded as distinguishable. 
Consider the order ABCD.  
This 2-d region of shape space is bounded by double collision line segments, 
and simultaneous double and triple collision points as in a).  
Next consider which regions lie adjacent to this, and continue until completion.  
b)  If reflection is quotiented out, ABCD and DCBA are indistinguishable etc.  
The 12 triangular regions, 18 line segments, 3 simultaneous double collision points and 
4 triple collision points can then be represented by b); note the topological identification.  
c) This may now be achieved by folding so as to join the ends of the bold line into a circle, 
followed by identifying the two edge circles to form $\Re P^2$ alongside the abovementioned line and point strata.  
d) If, instead, reflection is not quotiented out, there are then 24 triangular regions, 36 line segments, 
6 simultaneous double collision points and 8 triple collision points; note the topological 
identification. 
e) This may now be folded along the bold lines into a decorated tetrahaedron.  
Projecting this onto a circumscribing sphere represents each of the 2-d regions equally 
as isosceles spherical triangles with angles $\pi/2$, $\pi/3$, $\pi/3$, as expected from 
simultaneous double collision vertices having 4 emanating line segments 
while triple collision vertices have 6 emanating line segments.  
\normalsize}
\end{figure}
%
Note the additional choice of whether one quotients out the reflection.
Note that shape space has no place for a maximal collision.   
To construct corresponding RCS's, consider each shape space to be a `shell' to be 
crossed with the half-line (Fig 2c is an example).  
Then the maximal collision appears as the intersection of the lines built up from 
collision points on each `shell'.  
These mathematical observations form the basis of useful comparison with GR in Sec 4.3.  

Gergely and McKain \cite{GergelyMcKain} studied the $N = 3$, $d = 3$ case after Gergely 
\cite{Gergely} noted it was exceptionally well-behaved among the $d = 3$ models.  
Moreover, I note that the $N = 3$, $d = 3$ configuration space is classically indistinguishable from 
the $N = 3$, $d = 2$ one, leading me to conjecture that this good behaviour is more generally a 
result for 2-$d$ models.  
%
My study here of $d = 1$, $N = 3, 4$ for the BB82 RPM and $d = 1$, $N = 4$ for SIPT \cite{Albouy} 
complements these studies as regards laying out the classical structure of the simplest 
relational models possessing certain interesting features, including curved geometry, 
edges and singularities.

\subsection{Metrics and coordinate choices}

As a follow-on from {\bf I}, I start with $N = 3$, $d = 1$, 
both as it admits a nontrivial BB82 formulation and to serve as a guideline toward setting up 
the nontrivial $N = 4$, $d = 1$ SIPT model.  
From $N$ arbitrary, $d = 1$ relational Jacobi coordinates $R_i$, I pass to 
\be
\s_{\ba} \equiv \S_{\ba} = \frac{R_{\ba}}{R_{N - 1}}  \mbox{ } ,
\s_{N - 1} \equiv \N = R_{N - 1} \mbox{ } , 
\label{ratios}
\ee
i.e. a maximum set of independent shape coordinates $\S_{\ba}$ and one remaining non-shape coordinate $\N$.  
(\ref{ratios}) inverts trivially (away from $\N = R_{N - 1} = 0$).  
Then  
\be
\frac{2\st}{\mu} = \dot{R}_{a}\delta^{ab}\dot{R}_{b} = 
\dot{s}_c\frac{\pa R_a^{\Tr}}{\pa \s_c}\delta_{ab}\frac{\pa R_b}{\pa \s_d}\dot{s}_d \equiv 
\mbox{\sffamily G}^{ab}\dot{\s}_a\dot{\s}_b \mbox{ }.
\ee
for $\mbox{\sffamily G}^{ab}$ the RCS metric 
and $\mu$ the common mass by choice of normalization of the Jacobi coordinates.  
The conjugate momenta are 
\be
\left(
\stackrel{        p_{\sS}^{\ba}        }
         {        \mbox{\scriptsize $p$}_{\mbox{\tiny N}}          }
\right)^i 
= {p_{\ss}}^i = \mu \mbox{\sffamily G}^{ij}\dot{\s}_j \mbox{ }, 
\ee
while 
\be
\se = {\cal Q}(\S_{\bc},\N; {p_{\sS}}^{\bc}, p_{\sN}) = 
\frac{1}{2\mu}\sum_{a = 1}^{N - 1}\sum_{b = 1}^{N - 1}\mbox{\sffamily G}^{- 1}_{ab}p_{\ss}^ap_{\ss}^b + \sv 
\mbox{ } .
\label{Hcl}
\ee
I next note firstly that the $\q_I \longrightarrow \R_i$ map (Sec {\bf I}.4) 
has yet another useful property adapted to RPM's: 
it maps ${\cal E}$ in form-preserving fashion 
$\sum_{I = 1}^{N}\q_I\cdot\p^I$ to $\sum_{i = 1}^{N - 1}\R_i\cdot\P^i$. 
Secondly, $R_i \longrightarrow s_{\ba}$ maps $E$ to 
$\sum_{a = 1}^{N - 2}\S_{\ba}\N\frac{{p_{\tS}}^{\ba}}{\sN} + \N
\left( 
- \sum_{a = 1}^{N - 2}s_{\ba}\frac{{p_{\tS}}^{\ba}}{\sN} + p_{\sN}
\right) 
= \N p_{\sN}$.  
Thus in the SIPT model, away from the $\N = 0$ edge, $p_{\sN} = 0$, so (\ref{Hcl}) becomes 
\be
\se = {\cal Q}(\S_{\ba},\N; {p_{\sS}}^{\bb}) = 
\frac{1}{2\mu}\sum_{\ba = 1}^{N - 2}\sum_{\bb = 1}^{N - 2}\mbox{\sffamily C}_{\ba\bb}p_{\sS}^{\ba}p_{\sS}^{\bb} 
+ \sv(\S_{\bc}, \N) 
\mbox{ } , 
\label{Hclshape}
\ee
where $\mbox{\sffamily C}_{\ba\bb}$ is the shape metric 
obtained by striking out the end row and column of the inverse RCS metric.    
This metric plays an important r\^{o}le in the QM set up in Sec 5. 

For $N = 3$ (for which Jacobi coordinates are provided in Sec {\bf I}.6), 
the relational space metric is
\be
\mbox{\sffamily G}^{ab} = 
\left(
\stackrel{              \sN^{\mbox{\tiny 2}}          \mbox{ } \mbox{ } \mbox{ }      \sN\sS    }
         {              \sN\sS          \mbox{ }\mbox{ }       \mbox{\scriptsize 1} + \sS^{\mbox{\tiny 2}}    } 
\right)^{ab} \mbox{ } ,
\ee
while the shape space metric $\mbox{\sffamily C}$ is $1 \times 1$ and hence trivially flat.

I next set up $N = 4$.  
The relevant, relational subset of the arbitrary-$d$ {\it Jacobi K-coordinates} is 
\be
\R_1 = \q_1 - \q_2 
\mbox{ } , \mbox{ }
\R_2 = \q_3 - \frac{m_1\q_1 + m_2\q_2}{m_1 + m_2}
\mbox{ } , \mbox{ }
\R_3 = \q_4 - \frac{m_1\q_1 + m_2\q_2 + m_3\q_3}{m_1 + m_2 + m_3}
\mbox{ } ,
\ee
which (just as in {\bf I} for $N = 3$)  
I can take alongside any convenient absolute coordinate $\R_4 = \q_4$, say, and 
then normalize the coordinate definitions to be associated with a single mass $\mu$.   

Then the $N = 4$, $d = 1$ relational space metric is  
\be
\mbox{\sffamily G}^{ab} = 
\left( 
\stackrel{              \sN^{\mbox{\tiny 2}}  \mbox{ } \mbox{ } \mbox{ }    0    \mbox{ } \mbox{ } \mbox{ } \mbox{ } \mbox{ } \mbox{ } \mbox{ }  \sN\sS_{\mbox{\tiny 1}}        }
         {\stackrel{     0   \mbox{ } \mbox{ } \mbox{ }   \sN^{\mbox{\tiny 2}}   \mbox{ } \mbox{ } \mbox{ } \mbox{ } \mbox{ } \mbox{ } \mbox{ }  \sN\sS_{\mbox{\tiny 2}}        }
                   {  \mbox{ } \mbox{ } \sN\sS_{\mbox{\tiny 1}}   \mbox{ }  \mbox{ } \sN\sS_{\mbox{\tiny 2}} \mbox{ }     
    \mbox{\scriptsize 1} + \sS_{\mbox{\tiny 2}}^{\mbox{\tiny 2}} + \sS_{\mbox{\tiny 2}}^{\mbox{\tiny 2}}    }     }      
\right)^{ab} \mbox{ } ,  
\label{nontrss}
\ee
giving the shape space metric 
\be
\mbox{\sffamily C}_{\ba\bb} = 
\left( 
\stackrel{     \mbox{\scriptsize 1} + \sS_1^2  \mbox{ }   \sS_1\sS_2 }
         {\mbox{ } \sS_1\sS_2 \mbox{ } \mbox{\scriptsize 1} + \sS_2^2     }  
\right)_{\ba\bb} 
\label{snfl}
\ee
up to a factor of $\N^{-2}$.  
[N.B that in SIPT, as $\sv - \se$ is homogeneous of degree --2, all the $\N$ dependence of 
(\ref{Hcl}) forms a common factor, so that (\ref{Hcl}) is in effect free of $\N$.]  

For arbitrary $N$ {\sl and} $d$, I choose variables as follows.  
1 scale, $N$ - 1 ratios, $d(d - 1)/2$ absolute Euler angles 
and $N(d - 2) - d(d - 1)/2$ remaining angles of purely relational significance.  
E.g. for $N = 3$ and without loss of generality $d = 2$ and away from $\R_2 = 0$, 
\be
\S_1 = \frac{|\R_1|}{|\R_2|}\mbox{sin}\theta \mbox{ } , \mbox{ } 
\S_2 = \frac{|\R_1|}{|\R_2|}\mbox{cos}\theta \mbox{ } , \mbox{ } 
\N_x = R_{2x} \mbox{ } , \mbox{ }
\N_y = R_{2y}
\ee
are useful coordinates because they invert cleanly.  
These easily permit the now $4 \times 4$ semi-relational metric $\mbox{\sffamily G}^{ab}$ 
and the $2 \times 2$ shape metric  $\mbox{\sffamily C}_{\bar{a}\bar{b}}$ to be computed 
along the above lines (${\cal L}_{\alpha}$ and ${\cal E}$ now combining to jointly strike out the 
mixed scale/absolute direction coordinates' momenta $P_{\sN}^x$ and $P_{\sN}^y$).

\subsection{Edges and singularities of configuration spaces}

RCS is an analogue of superspace.  
RCS contains a better-behaved shape space (as it has no maximal collision) 
in a close parallel of how superspace contains a better-behaved conformal superspace (for example in the 
simple sense pointed out by DeWitt \cite{DeWitt67} of a 
$5 \times 5$ block of the $6 \times 6$ metric having better-behaved geometry).   
It is the configuration spaces without reflections quotiented out (corresponding to 
quotienting out the connected component of Eucl or Sim) that correspond to what 
Fischer and Moncrief \cite{FischMon} call `quantum' versions of superspace and conformal superspace.  
Perhaps the simpler study of RPM's will shed some light on the necessity and implications of 
studying `quantum' rather than `full' configuration spaces.  

A principal idea of the configuration space study approach 
is representing motions by geodesics on configuration space   
(e.g. just straight lines for $d = 1$, $N = 3$ free motion).  
As geodesics can lead into the edges of configuration space,  
(e.g. the $N = 3$, $d = 1$ straight (half-) lines that go into the triple collision), 
it is interesting to investigate the nature of these edges.
It is then beneficial that the study of singularities 
for particle mechanics is well-developed \cite{Diacu}, 
particularly for the classical $1/|\r_{IJ}|$ potentials 
but also for similar potentials such as $1/|\r_{IJ}|^l$.  
While studying this at the level of RCS and shape space 
is less common, at least some such studies have been done \cite{Montgomery2, Gergely, GergelyMcKain}.    
Ideally, one would like to know how typical it is for the motion to hit such a boundary 
and what happens to the motion after hitting the boundary.  
One possibility is that boundaries are singular, 
a simple subcase of this being when the boundaries represent curvature singularities 
of the configuration space.  
E.g. the  $N = 3$, $d = 2$ or $3$ RCS metric blows up at the 
triple collision \cite{GergelyMcKain}, while the shape metric is better behaved in being of  
constant positive curvature \cite{Montgomery2}.  
Also, the $N = 4$, $d = 1$ shape metric (\ref{snfl}) is of finite curvature.  

It should furthermore be added that one requires not the `bare' metrics 
$\mbox{\sffamily M}_{\Gamma\Delta}$ 
but conformally related 
$\tilde{\mbox{\sffamily M}}_{\Gamma\Delta} = 
(\se - \sv)\mbox{\sffamily M}_{\Gamma\Delta}$ for each $\se - \sv$ 
in order to encode motions as geodesics.  
Now, clearly, performing such a conformal transformation 
generally alters geodesics and curvature.   
Additionally, this transformation cannot necessarily be performed on extended regions 
since it requires a smooth nonzero conformal factor  
while $\se - \sv$ may have zeros or unbounded/rough behaviour.  
Thus more is required than a configuration space study by itself. 
The models in hand provide an interesting qualitative realization of this: 
geodesics can approach the triple collision from all directions in RCS  
while the dynamical trajectories of $1/|\r_{IJ}|$ potentials can only do so from 
the directions picked out by the central configurations (Euler solutions and Lagrange solutions 
for $N = 3$) \cite{Diacu}.  
Given such potential-dependent effects for RPM's, I ask whether the 
reflection condition proposed in the GR context \cite{DeWitt70, Misref} 
is really of general applicability.   

It is also important that models with ever-increasing $N$ are simple to envisage and represent, 
in contrast with how minisuperspace goes up to a maximum of 
3 degrees of freedom in vacuo or six with matter.  
Thus `stability of microspaces' is more easily studied for RPM's 
than for minisuperspace \cite{KucharRyan}.    
Moreover, for RPM's a number of important features and effects are known to require $N$ 
above certain values (see Sec {\bf I}.5 for discussion of noncollision singularities).  

Further QM aspects of the study of configuration spaces are touched upon in Sec 8.

\section{Relational space and shape space quantization}

Throughout I use the position representation, so $\hat{\q}_{A} = {\q}_{A}$ and 
$\hat{\p}_{A} = -i\hbar\pa/\pa{\q}_{A}$.  
I identify the approach of Sec {\bf I}.6--8 as the full- (for $d = 1$) or semi- (for $d > 1$) 
RCS analogue of superspace quantization \cite{Smolin}.  
I.e. ({\bf I}.34) says $\Psi = \Psi(r_{12}, r_{23} \mbox{ alone})$: 
translation invariance, just like (\ref{MOM}) says $\Phi = \Phi({\cal G}^{3} \mbox{ alone})$: 
dependence on 3-geometry and invariance of how coordinate lines are  
painted upon that geometry.  
I also note that constraining before quantizing 
and vice versa can both be done here and are equivalent.  

Likewise, the below approach to SIPT is a `shape space analogue' of conformal superspace quantization.  
The quantum constraints are then: the parts of ({\bf I}.34--36) required in dimension $d$,   
and a new one,   
\be
\hat{{\cal E}} \Psi(\q_A) \equiv \sum_{I = 1}^{N} \q_{I} \cdot \frac{\pa}{\pa \q_{I}} \Psi(\q_A) = 0 
\mbox{ } .  
\label{sippqc2}
\ee
I affirm that the commutator algebra of these four quantum constraints also closes, 
again just like the corresponding Poisson brackets, 
so that no further constraints appear at the quantum level.    
Note furthermore that the above is however no longer free of operator-ordering ambiguities: 
portions of the $q_A$'s could be on the other side of the $\pa/\pa q_A$'s.  
As written, which is in direct analogy with the ordering choice in the superspace quantization, 
({\bf I}.34) and (\ref{sippqc2}) say that one has not only $\Psi = \Psi$(relative coordinates alone), 
but also the Euler condition of homogeneity of degree 0, 
so $\Psi = \Psi$(ratios of relative coordinates alone).  
Moreover, there is a `conformal anomaly' in that each ordering appears to provide a 
distinct quantum theory.  For, while the condition remains an Euler homogeneity condition, 
its degree varies from order to order.   
Quantum algebra closure does not provide a selection principle to overcome this.  

As regards specific SIPT problems,  from (\ref{Hcl}) for $N = 3$, quantum-mechanically 
\be
\se\Psi   =   \frac{1}{\N^2}\left(
(1 + \S^2)    \frac{\pa^2 \Psi}{\pa \S^2} - 
2\N\S         \frac{\pa^2\Psi}{\pa \S\pa \N} + 
\N^2           \frac{\pa^2 \Psi}{\pa \N^2}
\right) 
+ \sv \Psi \mbox{ } .  
\ee
But, paralleling the argument of Sec {\bf I}.6, and using the working above (\ref{Hcl}), 
$\Psi$ depends on shape coordinates alone, while $\N$ is no true shape coordinate so 
\be
(\overline{\se} - \overline{\sv})\Psi = (1 + S^2)\frac{\d^2 \Psi}{\d \S^2} 
\mbox{ } ,
\ee
for the overlined quantities having factors of $\N^{-2}$ taken out.  
Repeating for $d = 1$, $N \geq 4$ leaves the less trivial quantum problem  
\be
(\overline{\se} - \overline{\sv})\Psi = 
\mbox{\sffamily C}_{\ba\bb}\frac{\pa^2 \Psi}{\pa \S_{\ba} \pa \S_{\bb}} = 
(1 + \S_1^2)  \frac{\pa^2\Psi}{\pa\S_1^2}           +
2S_1S_2       \frac{\pa^2\Psi}{\pa\S_1\pa\S_2}      +
(1 + \S_2^2)  \frac{\pa^2\Psi}{\pa\S_2^2}
\mbox{ } .
\ee
As argued in Sec 4.2, these are completely free of N, 
and hence are quantum problems of pure shape.  
These look like fairly normal TISE's, 
albeit they have the complications that they are on curved manifolds 
and have the rigid restriction that the potentials $\sv$ are homogeneous of degree $-2$.  
Nonseparability is prone to arise thus, 
so these problems are likely to require approximate or numerical investigation in a further paper.

\section{Internal time for the relational formulation of NM }

One place to seek for an internal time for RPM's is  
among the theories' natural scalars: $J$, ${\cal E}$, $\st$...
A restriction on internal time candidates is monotonicity.  
NM's Lagrange--Jacobi relation 
\be
\dot{\cal E} = \frac{\ddot{{J}}}{2} = 2\st + l\sv = 2\se + (l - 2)\sv
\ee
guarantees that the Euler quantity ${\cal E}$ is monotonic in many substantial cases (Fig 4).   
\begin{figure}[h]
\centerline{\def\epsfsize#1#2{0.45#1}\epsffile{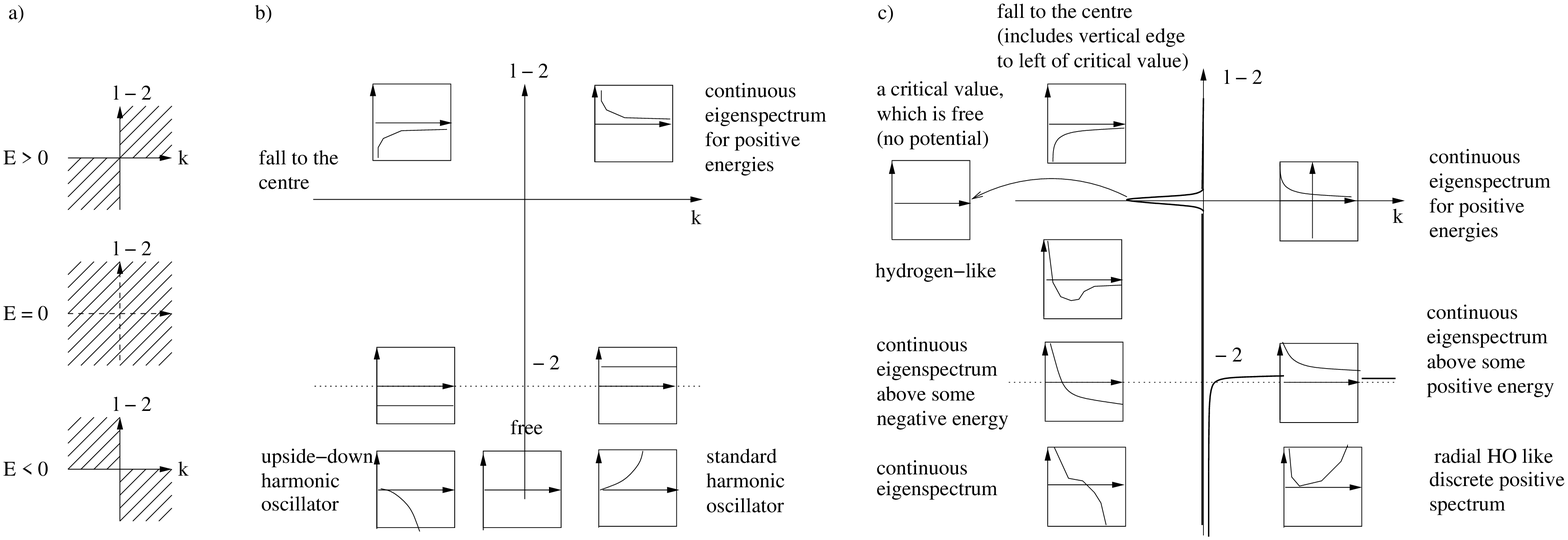}}
\caption[]{\label{FIG4.ps}
\footnotesize 
a) The shaded regions (dashed lines excluded) are the 
portions of NM with potential $\frac{k}{r^{l}}$ for which ${\cal E}$ is 
monotonic. 

For comparison, I provide diagrams b) and c) to illustrate how 
quantum behaviour varies over (k, l) parameter space for plain 1-d and 3-d 
radial effective potentials $\frac{k}{r^{l}} +\frac{\ml(\ml + 1)}{r^2}$ respectively.  
\normalsize}
\end{figure}
Furthermore, ${\cal E}$ is NM's analogue of GR's York time.    
Moving from SIPT to BB82 RPM recast as in this section 
is in close parallel to moving from maximal slicing to constant mean curvature-slicing in GR.  
In both, a frozen quantity is replaced by a monotonic time function.    
An even closer parallel is moving between the conformal gravity alternative theory in \cite{CG} 
and the `true degrees of freedom of GR' formulation in \cite{ABFKO}.  
In this case the additional parallels are: 
1) SIPT is scale invariant while conformal gravity is conformally invarianct.  
2) Both these invariances lead to homogeneity requirements in constructing actions for the theories.  
3) These are both taken care of by incorporation of powers of some dimensionful quantity: 
$J$ in SIPT and the volume of the universe in conformal gravity.  
4) This means that each action is expressible in terms of shape variables alone.  
5) In both cases the unfreezing process is tied to the incorporation of a global scale variable 
whereby such homogeneity is no longer necessary. 
 
For the $N = 3$, $d = 1$ BB82 RPM, I consider coordinates\fn{That $\sigma$ is 
a function of the (here flat relation space) metric and logarithmic should also be noted as 
counterparts of the properties of `Misner time' for minisuperspace.}
\be
s_1 \equiv S = f(R_1/R_2) \mbox{ } , \mbox{ a shape coordinate to be specially selected}
\label{bill}
\ee
\be
s_2 \equiv \sigma = \frac{1}{2}\mbox{ln}\left(R_1^2 + R_2^2\right) = \mbox{ln} {\sqrt {J/\mu}} 
\mbox{ } , \mbox{ a scale chosen strategically to be conjugate to ${\cal E}$ }
\label{spoon}
\ee
which invert to 
\be
R_1 = \pm \frac{f^{-1}(S)e^{\sigma}}{\sqrt{([f^{-1}(S)]^2 + 1)}}  
\mbox{ } \mbox{ }  , \mbox{ } \mbox{ }
R_2 = \pm  \frac{e^{\sigma}}{\sqrt{([f^{-1}(S)]^2 + 1)}} \mbox{ } . 
\label{king}
\ee
Then, for $^{\prime} = \frac{\pa}{\pa s_1}$,  
$$
\frac{2\st}{\mu} = \dot{R}_{a}\delta^{ab}\dot{R}_{b} = 
\dot{s}_c\frac{\pa R_a}{\pa s_c}^{\mbox{\scriptsize T}}\delta^{ab}\frac{\pa R_b}{\pa s_d}\dot{s}_d 
= \dot{s}_c\frac{e^{s_2}}{\sqrt{s_1^2 + 1}}
\left(
\stackrel{    \frac{(f^{-1})^{\prime}}{[f^{-1}]^2 + 1} \mbox{ } \mbox{ } -\frac{(f^{-1})^{\prime}f^{-1}}{[f^{-1}]^2 + 1}    }
         {    \mbox{\scriptsize $f^{-1}$}  \mbox{ } \mbox{ } \mbox{ } \mbox{ } \mbox{ }  \mbox{ }   
               \mbox{ }       \mbox{\scriptsize 1}                        }
\right)^{cb}
\dot{s}_d\frac{e^{s_2}}{\sqrt{s_1^2 + 1}}
{\left(
\stackrel{    \frac{    (f^{-1})^{\prime}    }{    [f^{-1}]^2 + 1    }      
              \mbox{ } \mbox{ } \mbox{ } \mbox{ } \mbox{ } \mbox{ }  
               f^{-1}    }
         {    \mbox{\scriptsize $-\frac{    (f^{-1})^{\prime}f^{-1}    }{    [f^{-1}]^2 + 1    }$} 
              \mbox{ } \mbox{ }   \mbox{\scriptsize 1}      }
\right)_b}^d
$$
\be
= e^{2s_2}
\left(
\stackrel{\frac{([f^{-1}]^{\prime})^2}{([f^{-1}]^2 + 1)^2}  \mbox{ }             0      }
         {  0  \mbox{ } \mbox{ } \mbox{ } \mbox{ } \mbox{ } \mbox{ } 
               \mbox{ } \mbox{ } \mbox{ } \mbox{ } \mbox{ }    \mbox{\scriptsize 1}      } 
\right)_{cd}
\dot{s}_c\dot{s}_d \equiv 
\mbox{\sffamily G}^{ab}\dot{s}_a\dot{s}_b 
\mbox{ } \Rightarrow \mbox{ } 
\st = \frac{\mu e^{2\sigma}}{2}(\dot{s}^2 + \dot{\sigma}^2)
\ee
if $f^{-1}$ obeys $([f^{-1}]^\prime)^2/([f^{-1}]^2 + 1)^2 = 1$, 
which it does without loss of usefulness if  
\be
S = \mbox{arctan}(R_1/R_2) 
\mbox{ } ,
\ee
then the shape space is flattened out and the whole RCS is conformally flat.  
Then the inverse transformation takes the form
\be
R_1 = e^{\sigma} \mbox{sin}S  \mbox{ } ,
\ee
\be
R_2 = e^{\sigma} \mbox{cos}S  \mbox{ } ,
\ee
so that as well as the aforementioned simplification, this choice of $S$ has an illuminating 
geometrical significance as the angle round the RCS of Fig 4d.

Then the conjugate momenta are 
\be
\left(
\stackrel{    P_{\sS}   }{    \mbox{\scriptsize $P_{\sigma}$}    }
\right)^i 
\equiv P_{\sS}^i = \mu \mbox{\sffamily C}^{ij}\dot{s}_j = \mu e^{-2\sigma}
\left(\stackrel{\dot{S}}{\dot{\sigma}}\right)^i 
\mbox{ } .
\ee
But in terms of the original coordinates, also 
\be
{P}_{\sigma} = \mu\dot{R}_1R_1 + \mu\dot{R}_2R_2 = \sum_{i = 1}^2 P^iR_i \equiv {\cal E} 
\mbox{ } .
\ee
I then note that $(Q, q_i; P, p_i) \longrightarrow (P, q_i; -Q, p_i)$ is straightforwardly a 
canonical transformation.  
Thus the whole of my move $(R_1, R_2; P_1, P_2) 
\longrightarrow (\sigma, S; {\cal E}, {P}_S)  \longrightarrow ({\cal E}, S; -\sigma, {P}_S)$
is a canonical transformation.  

I then regard  
\be
\se \equiv {\cal Q}({\cal E}, S; -\sigma, {P}_S) = \st + \sv = \frac{\mu e^{2\sigma}}{2}
\left[
\left(
\frac{    {P}_{\sigma}    }{    \mu e^{2\sigma}    }
\right)^2 + 
\left(
\frac{    {P}_S    }{    \mu e^{2\sigma}    }
\right)^2
\right] + \sv
= \frac{e^{-2\sigma}}{2\mu}({\cal E}^2 + P_S^2) + \sv(\sigma, S)
\ee
as an equation for $\sigma$, so that $\se = {\cal Q}$ is replaced by 
$P_{\epsilon} \equiv -\sigma = -\sigma({\cal E}, S; {P}_S) \equiv 
H^{\mbox{\scriptsize true}}({\cal E}, S; {P}_S)$, 
which upon quantizing in the new position representation, 
gives a TDSE 
\be
i\hbar\frac{\pa\Psi}{\pa \tau_{{\cal E}}} = \widehat{H}^{\mbox{\scriptsize true}}
\left(
\tau_{{\cal E}}, S, \hat{{P}}_S 
\right) 
{\Psi}
\ee
where I replace the symbol ${\cal E}$ by $\tau_{\cal E}$ as a token that 
the Euler quantity is playing the conceptual role of an internal time. 
Achieving this involves an algebraic problem which is explicitly soluble at least in some cases.
For example, for the $\se > 0$ free problem,  
\be
e^{2\sigma} = \frac{{\cal E}^2 + {P}_S^2}{2\mu\se} \mbox{ } ,
\ee
which gives a real logarithmic expression for $\sigma$ and then an explicit TDSE 
\be
i\hbar\frac{\pa\Psi}{\pa \tau_{{\cal E}}} = 
\frac{1}{2}
\left(
\mbox{ln}
\left(
\tau_{{\cal E}}^2 - \hbar^2\frac{\pa^2}{\pa S^2}
\right) 
- \mbox{ln}(2\mu\se)
\right)  
{\Psi}  
\mbox{ } .  
\ee
Another example is the $\se > 0$ repulsive harmonic oscillator $\sv = kR_1^2$, $k < 0$.  
Then in the new variables $\sv = k e^{\sigma} \mbox{sin}^2S $, so  
\be
e^{2\sigma} = \frac{    -\se \pm \sqrt{\se^2 + 2(-k)\mbox{sin}^2S(\tau_{{\cal E}}^2 +  P_S^2)}    }
                   {    2(-k)\mbox{sin}^2S    }
\mbox{ } . 
\ee
Then, with the given signs of $\se$ and $k$, the positive root option is  
required to give a real logarithm.  The subsequent explicit TDSE is   
\be
i\hbar\frac{\pa\Psi}{\pa \tau_{{\cal E}}} = 
\frac{1}{2}
\left( 
\mbox{ln}
\left(
\sqrt{ \se^2 + 2(-k)\mbox{sin}^2S
\left(
\tau_{{\cal E}}^2 -  \hbar^2\frac{\pa^2}{\pa S^2} 
\right)  
                                   } - \se
\right)
- \mbox{ln}(2(-k)\mbox{sin}^2S)
\right)
{\Psi} \mbox{ } .  
\ee
These examples illustrate typical features among the wider set of working examples that I have 
found (to be presented elsewhere \cite{Soland}).  
Namely, the true Hamiltonians contain roots and logs 
within which there is good positivity, as well as corresponding to nonconservative dynamics 
through their dependence on the internal time.  

Note the parallel between $J = e^{2\sigma} > 0$ here 
(as it's a sum of masses times squares) 
and $\psi > 0$ in the study of the Lichnerowicz--York type equation 
(required for it to be a mathematically meaningful conformal factor).  
While some of the Lichnerowicz--York type equation's difficulties stem from its elliptic operator LHS, 
more stem from it having a complicated polynomial RHS.  
For the analogue relational problem, there is no differential operator, but there is still  
complicated polynomiality: explicit solution is not possible here in some cases, 
if the polynomial is general and of high enough order 
(which is tied to the form of the potential function).  
%

%
What I show here is that this method extends to $N$ particles in 1-$d$. 
Choose as coordinates  
\be
\sigma = \mbox{ln}\sqrt{J/\mu} = 
\frac{1}{2}\mbox{ln}
\left( 
\sum_{i = 1}^{N - 1} R_i^2
\right)
\mbox{ strategically chosen to be conjugate to ${\cal E}$} 
\ee
and the $N - 2$ ratios of (\ref{ratios}).\fn{Now these can't be beaten flat.  
Further algebraic choices of coordinates to simplify the problem I leave for the future.} 
Then 
\be
\st = \frac{\mu e^{2\sigma}}{2}
\left(
\dot{\sigma^2} + \frac{ \Delta {\dot s}_{\ba}^2 - s_{\ba}\dot{s}_{\ba} s_{\bb}\dot{s}_{\bb} }
                      {\Delta^2}
\right)
\ee
for $\Delta \equiv \sum_{\bar{a}} s_{\bar{a}} \dot{s}_{\bar{a}}$ and  
\be
{P}_{\sigma} = \mu e^{2\sigma}\dot{\sigma} = \sum_{i = 1}^{N -1}P^iR_i \equiv {\cal E} 
\mbox{ } ,
\ee
\be
{P}_{S}^{\ba} = \mu e^{2\sigma}(\mbox{\sffamily C}^{-1})^{\ba\bb}\dot{S}_{\bb} 
\mbox{ } \mbox{ for } \mbox{ } 
(\mbox{\sffamily C}^{-1})^{\ba\bb} 
= \frac{\Delta \delta^{\ba\bb} - s^{\ba}s^{\bb}}{\Delta^2} 
\mbox{ } \mbox{ } , \mbox{ } \mbox{ (inverse shape space metric) .} 
\ee 
Invertibility of this is guaranteed at least as far as $N = 5$ (for which  
$\mbox{det}\mbox{\sffamily C} = \Delta^{N - 3} \neq 0$).    
Thus I can write 
\be
\se = {\cal Q}({\cal E}, S_{\ba}; - \sigma, {P}_{S\ba}) \equiv \frac{e^{-2\sigma}}{2\mu}
\left(
{\cal E}^2 + ({\cal C}_{\ba\bb})^{-1}{P}_{S}^{\ba}{P}_{S}^{\bb}
\right) 
+ V(J, S_{\bc}) 
\label{snark}
\ee
In any case, there is no $\sigma$ dependence within the first bracket in (\ref{snark}), so the 
$\sigma$-eliminating algebra for the $N = 3$ 
examples given above extends immediately to arbitrary $N$.  

Finally, I consider the $d = 2$ extension of $N = 3$.  
Now select 2 Sim-invariants $S_1$ and $S_2$ 
alongside the scale  
\be
\sigma = \mbox{ln}\sqrt{J/\mu} = 
\frac{1}{2}\mbox{ln}
\left( 
\sum_{i = 1}^{2} |\R_i|^2
\right)
\ee
and one independent coordinate N.  
As $S_1$, $S_2$ and $\sigma$ are Rot-invariant and $N$ is not (through its independence), 
${\cal L} = 0$ acts to fix $p_{\sN} = 0$ except possibly at configuration space boundaries.  

Thus I have found substantial portions of NM stripped of its absolute structure 
nevertheless possesses an internal time.  This should serve as a warning to those 
that consider privileged notions of time to be unnatural in relational theories (including GR) 
on aesthetic or philosophical grounds: {\sl theories that are relational can nevertheless 
provide a privileged notion of time from within their own structure.}  I furthermore 
mention some possible technical applications of this toy model for internal time in the conclusion.

\section{Semiclassical approaches}

This section's approach,
in which dynamics is contended to emerge for subsystems within stationary universes, 
is more widely applicable than the approach of Sec 6.    
Here, I use RPM's to investigate the semiclassical approach 
and objections raised against it in the literature.     
Of these, the two Barbour--Smolin objections \cite{BS} used RPM to suggest 
faults with the semiclassical approach in general, 
which is one reason this was out of favour as an explanation 
in Barbour's works \cite{B94II, EOT}.  
But I got around these objections in {\bf I} for the BB82 RPM,\fn{I comment here on 
the parallel situation in SIPT.  
The sensitivity of spread to small masses objection was resolved for BB82 RPM 
by the standardness of the underlying formal mathematics 
and by adhering to the relational interpretation 
whereby spread is in relative separation (and relative angle).  
For SIPT, the corresponding interpretation is in terms of {\sl spreads in relational ratios} and 
moreover that now the underlying formal mathematics is nonstandard (at least as far as I 
know and in the context of the QM literature).  
This nonstandardness then makes SIPT small universe models somewhat harder to handle.  
But as I am currently contemplating SIPT purely as a toy for geometrodynamics rather than as a 
realistic particle mechanics, I do not attempt a `recovery of reality' for it.} 
enabling me now to consider how much explanatory power its semiclassical approach does afford, 
alongside the line of thought of consistent records approaches that Barbour has favoured instead.   

For BB82 RPM, consider splitting the relative Jacobi coordinates $\R_i$ into 
heavy and light coordinates $\uH_{\jp}$ and $\uL_{\kpp}$, 
associated with cluster reduced masses of order of magnitude 
$M_{\sH} >> M_{\sL}$.   
This is possible e.g. for 1, 2 H particles of similar mass $M$ and 3 an L particle of mass $m_3$, 
whereupon $\R_1$ is a heavy Jacobi coordinate $\R_{\sH}$ 
and $\R_2$ is a light Jacobi coordinate $\R_{\sL}$, 
\be
M_{\sH} \approx M/2 << m_3 = M_{\sL} \mbox{ } .
\label{Po}
\ee
In general this heavy-light split is distinct from the shape--scale split hitherto used 
in this article.\fn{But the two can sometimes be aligned \cite{Soland},   
corresponding to shape-changing motions dominating or being dominated by 
scale changing motions.  This is of relevance to cosmology, while shape also plays 
an important role in $N$-body problem studies (see e.g. \cite{Diacu}).}  
 
Consider systems with energy constraint taking the form 
\be
\hat{{\cal Q}}\Psi = 
\left(
\hat{{\cal Q}}_{\sH}(\uH_{\jp}) + \hat{{\cal Q}}_{\sL}(\uH_{\jp}, \uL_{\kpp}) 
\right)
\Psi = \se\Psi 
\mbox{ } ,
\ee
\be
\hat{{\cal Q}}_{\sH} = 
-\frac{\hbar^2}{2M_{\sH}} \sum_{\ip} \frac{\pa^2}{\pa \uH_{\ip}^2} +  \sv(\uH_{\jp}) 
\mbox{ } . 
\ee  
Consider the wavefunction to be of the `extended WKB form' 
\be
\Psi = e^{iM_{\tH}\ssW\left(\usH_{\ip}\right)/\hbar}\psi(\uH_{\jp}, \uL_{\kpp})
\mbox{ } .
\label{WKB}
\ee  
Then by the Leibniz rule and cancelling off the common exponential factor,
\be
\hat{{\cal Q}}_{\sL}\psi + \sv\psi - \frac{\hbar^2}{2M_{\sH}}\sum_{\ip}
\left(
-\frac{M_{\sH}^2}{\hbar^2}
\left(
\frac{\pa \sW}{\pa \uH_{\ip}}
\right)^{2} 
+ \frac{iM_{\sH}}{\hbar}\frac{\pa^2 \sW}{\pa \uH_{\ip}^2} 
+ \frac{2iM_{\sH}}{\hbar}\frac{\pa \sW}{\pa \uH_{\ip}} \cdot \frac{\pa}{\pa \uH_{\ip}} 
+ \frac{\pa^2}{\pa \uH_{\ip}^2}
\right)
\psi
 = \se\psi
\label{creepers}
\ee
the first approximation to which is the Hamilton--Jacobi equation
\be
\frac{M_{\sH}}{2} \sum_{\ip} 
\left(
\frac{\pa \sW}{\pa \uH_i}
\right)^2
+ \sv - \se = 0 \mbox{ } .
\ee
Of what remains
the semiclassicial approach stance is \cite{SCB, Banks},\fn{This involves a generalization of the 
Born--Oppenheimer procedure.}  to retain the cross-term 
while regarding the double derivative terms to have a negligible effect.  
Rearranging within the cross-term, 
\be
\sum_{\ip}
\frac{\pa \sW}{\pa \uH_{\ip}} \cdot \frac{\pa \psi}{\pa \uH_{\ip}} = 
\sum_{\ip}
\frac{    \P_{\sH}^{\ip}    }{M_{\sH}} \cdot \frac{\pa \psi}{\pa \uH_{\ip}}
\label{jeepers}
\ee
by a standard move of Hamilton--Jacobi theory.  

Next, differences begin to appear between the RPM's 
and previously-studied absolute particle models \cite{Banks, BriggsRost}:    
the momentum-velocity formulae of RPM's are complicated by auxiliary correction terms.    
Moreover, the way I am approaching the problem, 
the exceedingly simple translational auxiliaries have been eliminated a priori 
by using relative rather than particle position coordinates.   
This approach shows that 1-d BB82 RPM models represent an insufficient modelling 
improvement in this respect (running against \cite{BS}, whose examples are restricted to 1-d models).  
For, in relative Jacobi coordinates, these look just like their spatially-absolute counterparts 
in particle position coordinates 
Then the momentum--velocity relations remain of the form\fn{It is because I treat the models as parametrized 
that label-time $\lambda$ and `lapse' $\dot{O} = \pa t/\pa \lambda$ appear at this stage.} 
\be
\P_{\sH}^{\ip} = \frac{M_{\sH}}{\dot{O}}\frac{\pa \uH_{\ip}}{\pa \lambda} 
\mbox{ } ,
\ee
then by the chain-rule (\ref{jeepers}) becomes
\be
\frac{1}{\dot{O}} \sum_{\ip} 
\frac{\pa \uH_{\ip}}{\pa \lambda} \cdot \frac{\pa \psi}{\pa \uH_{\ip}} 
= \frac{1}{\dot{O}} \frac{\pa \psi}{\pa \lambda}
\ee
and what remains of (\ref{creepers}) is 
\be
i\hbar\dot{\psi} = \dot{O}\hat{{\cal Q}}_{\sL}(\uH_{\ip}, \uL_{\kpp}) \psi \mbox{ } .
\ee
  
A closer parallel with full geometrodynamics is struck 
by studying $d > 1$ RPM's in relative Jacobi coordinates.    
E.g.\fn{The $d = 2$ version of BB82 RPM and SIPT for $d$ = 1,2,3 
work analogously in this respect.} for $d = 3$ the momentum--velocity relations are then 
\be
\P_{\sH}^{\ip} = \frac{M_{\sH}}{\dot{O}}
\left(
\frac{\pa\uH^{\ip}}{\pa \lambda} - 
\frac{\pa\underline{b}}{\pa\lambda} \crr \uH_{\ip}
\right)
\ee
so (\ref{jeepers}) becomes 
\be
\frac{1}{\dot{O}}
\left(  
\sum_{\ip} \frac{\pa \uH_{\ip}}{\pa \lambda} \cdot \frac{\pa \psi}{\pa \uH_{\ip}} - 
\sum_{\ip} \frac{\pa \underline{b}}{\pa \lambda} \crr \uH_{\ip} \cdot \frac{\pa \psi}{\pa \uH_{\ip}}
\right)
=
\frac{1}{\dot{O}}
\left(  
\frac{\pa \psi}{\pa \lambda} - 
\frac{\pa \underline{b}}{\pa \lambda} \cdot \sum_{\ip}  \uH_{\ip} \crr \frac{\pa \psi}{\pa \uH_{\ip}}
\right) \mbox{ } .  
\ee
But the zero AM constraint is separated in relative Jacobi coordinates ({\bf I}.39) so 
$\uH_{\ip} \cr \frac{\pa \psi}{\pa \usH_{\ip}} = 
\hat{\underline{L}}_{\sH}\psi = - 
\hat{\underline{L}}_{\sL}\psi $
and so the remainder of (\ref{creepers}) becomes 
\be
i\hbar \dot{\psi} = 
\left(
\dot{O}\hat{{\cal Q}}_{\sL}(\uH_{\ip}, \uL_{\kpp}) - \dot{\underline{b}}\cdot\hat{L}_{\sL}
\right) 
\psi \mbox{ } .
\label{Ypres}
\ee
Now note the strong parallel between (\ref{Ypres}) and the Tomonaga--Schwinger WKB-TDSE 
arising for minimally-coupled matter fields in GR under the approximation of a known fixed 
background.\fn{The RPMs' explicit eliminability permits one further approach: 
trading the above constraint contribution for a more complicated kinetic term.}

This approach depends on $\psi$ has to have some minor 
$\uH_{\ip}$ dependence (via ${\cal Q}_{\sL}$ having some residual 
and nonseparable functional dependence on the $\uH_{\ip}$), 
in order for the crucial chroniferous cross-term be nonzero.  
This does not cause a conceptual problem in practise -- 
a minimum guarantee of everything in the universe being coupled to everything else 
is generically 
guaranteed by the non-sheildability of gravity --    
but it does require further work beyond the scope of paper {\bf I}'s separable solutions to get 
to grips with this.  

{\sl If one accepts the assumptions of this Section},  
the emergence of quantum dynamics from a stationary state is explained.    
Then one can set up a stationary state universe consisting of a small 
subsystem (e.g. an atomic model plus a nearby grazing particle)  
in weak coupling with a large complement subsystem (e.g. a distant massive body), 
for which the small subsystem is capable of having a semblance of dynamics 
(with respect to a time provided by the large complement subsystem).  
If the small system is itself macroscopic, such as Barbour's kingfisher \cite{EOT},  
then the above reasoning would permit the semblance of flying if a large 
complement subsystem such as the Earth is present, and, moreover, if   
one also subscribes to the validity of `Ehrenfest manipulations', to {\it be} 
an overwhelmingly classical bird-object.  
The next section explores the whether this section's assumptions are plausible.

\section{Semiclassical critique and other timeless approaches}

In conventional QM, one presupposes that the quantum subsystem under study  
is immersed in a classical world, crucial parts of which are the observers and/or measuring apparatus.  
In familiar situations, NM turns out to give an excellent approximation for this classical world.   
It is the conventional absolutist conceptualization of NM 
that provides the external absolute time with respect to which the quantum system's dynamics occurs.  
However, there are notable conceptual flaws with extending this `Copenhagen' 
approach to the whole universe.    
For, observers/measuring apparatus are themselves quantum mechanical, 
and are always coupled at some level to the quantum subsystem. 
Treating them as such requires further observers/measuring apparatus so the situation repeats itself.  
But this clearly breaks down once the whole universe is included 
[by then the assumption of NM and the objective existence of something which in familiar situations 
passes for absolute time has also become unreasonable].  
N.B. (as detailed below) that some approaches to QM may be more useful than others when it comes to 
making this generalization.  
It is this in this more general whole-universe context that Barbour and I are asking questions 
about `simple subsystems'

\noindent such as  kingfishers and atoms.   

The above semiclassical scheme represents an {\sl attempt} 
to understand the replacement of the standard scheme with a TISE.  
This is in parallel with the GR Wheeler--DeWitt equation being a TISE, while avoiding the  
{\sl severe technical complications} that occur in the generic GR context.\fn{These include   
H -- L splits now being ambiguous, 
and the effect of higher order corrections on the conceptual structure.} 
Barbour did not follow the semiclassical approach, 
in part because of the two BS objections, but also 
in part because of a further three objections to using the WKB ansatz \cite{BS, B93}.  
These relate to two of its notable features and one associated interpretational issue, 
and are covered in the next three subsections.

\subsection{Is it justifiable for the ansatz to be complex? 
            Indeed, where do the complex numbers come from in QM ? }  

This was one issue Barbour raised in \cite{B93}.  
One answer I found to this in the literature (Landau and Lifschitz \cite{Landau}) 
is that complex numbers are brought into QM 
precisely by presupposing a WKB ansatz for the wavefunction.  
This leads straight to the problems in Sec 8.2.
Another answer (Dirac \cite{DiracQM}), however, is 
that complex numbers arise as a consequence of the following inadequacy of the real numbers.  
Conventional QM, at least, requires noncommuting operators.  
But consider 2 noncommuting real operators $\eta$, $\xi$.  
Then, $\eta\xi = \bar{\eta}\bar{\xi} = \bar{\xi\eta} \neq \xi\eta$.  
Therefore the product of $\eta$ and $\xi$ cannot be real.  
It follows from this that {\sl one requires a larger field 
than the reals over which to do QM}, and adopting the complex numbers for this purpose 
{\sl turns out to be an adequate choice}.\fn{An adequate choice (see \cite{B93, Schwinger})  
means any field possessing a structure J such that $\mbox{J}^2 = -$(identity).}  

Dirac's point of view provides a strong argument for the relevance of working over complex rather than 
real numbers in treating constrained closed universe model TISE's such as those of Sec 5,7 or the 
Wheeler--DeWitt equation.   
It is then natural to bring in complex wavefunction ans\"{a}tze for these.\fn{Barbour, 
however, is open to the possibility of quantum cosmology 
having {\sl substantially} different machinery and properties from ordinary QM.  
Were it technically demonstrated that this difference can run as deep as 
quantum cosmology not retaining the characteristic noncommutativity of QM, 
then my above extrapolation of Dirac's argument to quantum cosmology would cease to suffice.}

\subsection{Is it justifiable for the ansatz to contain a single piece?}

A wavefunction that is more general than the WKB ansatz (\ref{WKB}) 
is the superposition of a number of such terms \cite{BS, B93, Zehbook}.  
Furthermore, the working by which the semiclassical approach yields its TDSE 
ceases to work under this generalization.  
On these grounds, adopting this ansatz requires justification.  
While in ordinary QM it is an obvious justification that the WKB ansatz for the subsystem under study 
follows from the presupposition of immersion in a classical world,   
this clearly becomes an unacceptable presupposition in the extrapolation to the whole universe.  
Thus, the semiclassical approach to quantum cosmology relies on the WKB ansatz, 
and is in difficulty unless this ansatz can be justified independently of such 
`Copenhagen' vestiges.\fn{Along these lines, it is more satisfactory to set out to 
explain the classicality of present-day spacetime by quantum cosmology rather than considering this 
to be a `final condition' restriction on quantum cosmological solutions.}
While hopes have been expressed that the WKB ansatz will be independently justified by 
{\sl decoherence} (or otherwise), these hopes come with reservations 
(e.g. \cite{Kuchar92, POTlit2, SCAII, TH} are between far from and not entirely optimistic about this).   
Papers {\bf I}, {\bf II} serve to open up RPM's as one interesting and suitable arena 
wherein issues such as untainted origins for the WKB ansatz can be explored.

\subsection{\bf If a single WKB piece, why the particular sign?}

It is worth commenting further that Barbour \cite{B93, B94II} and previously Zeh \cite{Zeh} 
have pointed out the meaninglessness of asserting that a plus sign in the WKB exponent corresponds 
to an expanding universe and a minus sign corresponds to a contracting one.  
This plays a part in the debate of the mysteriousness of the Arrow of Time, 
and is related to the choice of boundary conditions postulated for the universe.    
BB82 RPM is a reasonable arena to investigate this (with increases and decreases in moment of inertia 
deputizing for the increases and decreases of the volume of the universe in GR).

\subsection{\bf Consistent Records Alternatives}

There has been a substantial shift in papers {\bf I} and {\bf II} from the rationale against 
the semiclassical approach in Barbour's works \cite{BS, B93, B94II, EOT}.  
Barbour and Smolin's objections specific to RPM's were overcome in 
in {\bf I}, while this paper overcomes Barbour's complex number objection.    
But the WKB ansatz and its properties 
remain unjustified 

\noindent in the context of whole universes.  
Thus a technically clearer rendition of whether the semblance of flight of kingfishers 
in \cite{EOT} is mysterious is that, 
within a closed universe described by a TISE, 
the semblance of flying of kingfishers {\sl is} mysterious 
{\sl if} one does not accept the imposition of the WKB ansatz, for which neither 
the usual QM justification holds, nor has an alternative justification been demonstrated.  

Thus, it is worth considering also more radical timeless approaches 
along the lines of consistent records formulations.   
Previous conceptual work in this field has been done by Page \cite{Page} and Barbour 
\cite{Bararr, B94II, EOT}, while Gell-Mann and Hartle \cite{GMH, Hartle} and 
Halliwell, Thorwart and Dodd \cite{H99, H00, TH, HD} have provided particular examples.   
Furthermore, the possibility that the semiclassical approach itself (or elements thereof) 
is required to account for these approaches should not be discounted.    
While, thinking in the opposite direction, it would be conceptually pleasing for histories approaches 
to be replaced by consistent records approaches 
(from which consistent histories can now be regarded as constructed rather than reconstructed).

\subsection{\bf Consistent records/time capsules in model particle universes}   

The following simple, non-relational example of Halliwell \cite{H99} 
suffices to obtain a clear, mathematical notion of record that is 
adaptable for use in RPM's.  
Consider a heavy particle moving through a medium of light particles.  
The heavy particle disturbs these into motion.  
Subsequent instants consist of the particles' postions and momenta.  
It is these instants which are the records, 
and the motion or history of the large particle can then be reconstructed 
(perhaps to some approximation) from them.   
It should be noted that this set-up requires not only a distinction between L and 
H variables but also the complication of these having intercoupled potentials.   
However, the simplifying feature that the sucessful modelling of the semiclassical effect 
does not require a populous environment of L particles 
is suggested by two of Halliwell's examples in \cite{H99}.    
This notion of record can be adapted to e.g. a $N = 3$, $d = 1$ BB82 RPM  
setting as follows.    
Consider the 2 H and 1 L particle situation of (\ref{Po}).  
In relational terms, this situation is the motion of  
a H and an L relative Jacobi separations.   
If these have coupled potentials 
(whether `by hand' with an illustrative linear interaction term 
or more realistically by a physical interaction such as gravitation), 
then the `heavy separation' will disturb the `light separation' into motion.  
Subsequent instants consist of inter-particle (cluster) separations and their momenta  
with respect to label time.  

While Halliwell's model permits the recovery of past history through a decoherence process, 
it is not so clear what sort of counterpart this may have 
in the (particularly time-) relational setting.  
N.B. to date only a handful of models \cite{H99, HD} 
have been cast in records formulation, all of which have external time.  
With GR in mind, this is an important deficiency to resolve, 
and the RPM's look to be a useful setting for doing this 
(albeit the need for coupling terms in the potential 
places this issue just beyond the scope of the current paper, while \cite{Soland} 
is being written to include this feature).  

Barbour's own approach \cite{Bararr, B94II, EOT} (may) differ in two respects.  
His `time capsule' vision of records come as part of a `heap' of instants.  
So, which characteristics distinguish time capsules from just any instants?  
Do these features amount to precisely the same as the above notion of consistent records? 
Is the semblance of dynamics/history in this scheme to be recovered by the same decoherence 
mechanism as in Halliwell's scheme?  
How come consistent records certainly appear to predominate in nature over the other 
sorts of instants in the `heap' (which would seem to be more numerous)?
As regards this last issue, Barbour speculates that this is enforced 
by the asymmetry of the geometry of the configuration space causing the wavefunction of the 
universe to concentrate on `time capsules'.  
%

\subsection{\bf Does RCS geometry drastically affect the distribution of $\Psi$ ? }

While Barbour \cite{B94II, EOT} has placed much emphasis 
on it being the reduced configuration space geometry that influences the distribution of the wavefunction,   
I justify here a somewhat different emphasis.   
It is well known that the form of the potential $\sv$ and the value of the energy $\se$ in general  
play an important part in determining the distribution of the wavefunction.  
This remains the case for RPM's, but for these one has the somewhat less usual situation that 
the curved, stratified reduced configuration space geometries play a role {\sl alongside} 
$\sv$ and $\se$ in determining the distribution of the wavefunction.  
Furthermore, one cannot in general redefine problems so that only geometry 
or only energy and potential affect the distribution of the wavefunction.  
For, $\se - \sv$ is a single function while configuration space geometry in general 
requires a number of independent functions to encode.  
Nor can $\se - \sv$ always be incorporated into the geometry by being considered to be a conformal 
factor, as the range of this transformation may not cover the entire physical regime of interest.

My shift of emphasis above is particularly justified as the simplest RPM studies 
(e.g. those in {\bf I}) have trivial reduced configuration space geometry, 
but nevertheless can have sharply-peaked weavefunctions.  
On the other hand, a major goal of this paper is to develop RPM's with nontrivial    
RCS or shape space, wherein geometry is expected to be a competitor in determining 
the distribution of the wavefunction.   
Toward such a study, I argue preliminarily from the nature of the approximations 
used in simple quantum chemistry calculations 
(of reasonable success in comparison with observation), 
that it is likely that deleting points or including edges on configuration space 
will not {\sl drastically} alter the distribution of the wavefunction over it.  
 
It is worth mentioning that both configuration space geometry 
and potential in general play a role in minisuperspace \cite{minimetric}.  
Moreover, there is an important technical distinction between RPM's and minisuperspace 
as regards configuration space geometry, 
since minisupermetrics are semi-Riemannian 
while RCS and shape space metrics are Riemannian.  
%
As conformal superspace is Riemannian-signature, building up an understanding of both situations is interesting.  
The technical distinction translates to solving both 
hyperbolic and elliptic TISE's on curved spaces.   
As regards elliptic equations, these are a class of problem known to be {\sl capable} of producing 
some sort of pattern which reflects the underlying shape \cite{Spots, Than}.
It still remains to be understood, however, whether the concentration of the wavefunction 
on such regions furthermore favours the selection of `time capsules'.

\section{Conclusion}

In this paper, I have set up and used the Barbour--Bertotti 1982 (BB82) and scale-invariant particle 
theory (SIPT) relational particle models (RPM's)   
as an arena for testing ideas about the canonical quantization of GR.  
I have covered 
thin sandwiches,  
the classical and quantum study based on configuration spaces, 
internal time, 
semiclassical approaches 
and timeless approaches based on records/time capsules. 
The account below covers both this paper's results 
and some remaining open questions.

The toy models of this paper afford not a thin sandwich conjecture like GR, 
but explicit thin sandwich eliminations.  
An open question then is: do these lead to path integral or histories approaches, 
thus vindicating Wheeler's underlying QM transition amplitude analogy?  

As regards study of configuration spaces, this paper 
has studied one of the simplest (shape space for 4 particles on a line), 
which complements Gergely's study of another of the simplest (RCS for $N \geq 3$ particles).  
These studies open up classical investigation of geodesics and singularities 
on these configuration spaces in parallel with existing programs in GR.  
One issue is the extent to which shape space is better-behaved than RCS (in parallel 
with conformal superspace being better-behaved than superspace).  Another is the effect on the 
quantum theory of quotienting out discrete transformations (here, reflections).   
Finally, QM issues can be investigated: 
the distribution of the wavefunction over the configuration spaces 
and what boundary conditions it satisfies.  
While DeWitt and Barbour in separate contexts advocate various 
aspects of the usefulness of studying the configuration space geometry, 
there are limitations on this in practice through the potential also playing a role.   
Features of this [e.g. how $1/|\r_{IJ}|$ potentials affect the 
direction of approach to the triple collision] could be studied in an RPM context. 
It would furthermore be useful to study 
how these effects change, still within an o.d.e setting, when one considers metric indefiniteness 
and the particular potentials of minisuperspace.  

I have identified the procedure performed in Sec {\bf I}.6--8 as full- and semi-RCS quantizations.   
I have set up shape space quantizations in this paper.  
These lead to time-independent Schr\"{o}dinger equations (TISE's) 
containing a curved configuration space metric.

I have found that one portion of Newtonian mechanics stripped of its absolute structure 
nevertheless possesses an internal time: the Euler internal time.  
I have provided time-dependent Schr\"{o}dinger equations with respect to this internal time 
for two choices of potential.  
It is furthermore worth investigating whether the Euler internal time--York internal time analogy 
is close enough to be a useful arena for investigation e.g. of global issues \cite{Kuchar92, Torre} 
and of Wang's alternative \cite{Wang} to York's procedure.

I have considered the semiclassical approach to BB82 RPM.   
I overcame two objections \cite{BS} specific to this in {\bf I}.  
I have presented a counter-argument to the further general objection that 
this ansatz is complex while the TISE is real.  
If the WKB ansatz for the wavefunction of the universe is accepted, 
there are no problems in accounting for a universe governed by a TISE giving rise to apparent 
dynamics for subsystems.  
But genuine closed-universe reasons for accepting the finer detail of this ansatz 
are still lacking, which affects both the above recovery of dynamics and one approach  
to resolving the arrow of time puzzle.  

This lack motivates the search for alternative explanations for 
the semblance of dynamics within closed universes.
To this end, I have extended Halliwell's simple but precise mathematical notion of records to RPM's.  
Barbour has a similar (but possibly in detail distinct) notion of `time capsules'.  
Past history/dynamics would then be constructs emergent from these notions of present instants.  
Such an approach would need to explain how records/time capsules strongly predominate 
among the universe's nows that we (appear to) experience.  
Barbour conjectures that this is due to `time capsules' being selected by the asymmetric shape of 
the configuration space causing the wavefunction to concentrate on these.  
I have provided restricted preliminary evidence that this conjectured mechanism 
is at most only part of the picture.  
Halliwell conjectures a decoherence explanation.  
It would be interesting to know if these are two parts of the same picture or incompatible.

Wider relations among these various approaches and notions of time are also of interest 
(see e.g. \cite{SCAII, Kuchar92,POTlit2}).    
While the semiclassical emergent WKB time approach is free of the portion restriction of 
the Euler internal time approach, can (or must)
these two notions of time be aligned for this portion?

The various Schr\"{o}dinger equations mentioned above --and variants-- merit further study:  
ordering ambiguities, spectral well-definedness, and eventual solution.  

Finally, further study of approaches involving semiclassical or decoherence elements 
requires harder, coupled-potential, nonseparable RPM's.  
While such problems were mentioned/set up in papers {\bf I} and {\bf II} 
(multi-Coulomb problems and shape space problems set up in Sec 5),
I leave carrying out approximate/numerical treatments for these 
for subsequent papers.  


\mbox{ }

\noindent{\bf Acknowledgments}

\mbox{ }

\noindent I thank Professor Don Page and Dr. Julian Barbour for many discussions. I also thank 
Professor Robert Bartnik, Mr. Brendan Foster, Dr Laszlo Gergely, Professor Gary Gibbons, 
Dr. Bryan Kelleher, Ms. Isabelle Herbauts, Professor Theodore Jacobson, 
Professor Jacek Klinowski, Professor Malcolm MacCallum, Professor Niall \'{O} Murchadha, 
Dr. Oliver Pooley, Professor Reza Tavakol, Professor Lee Smolin, Dr. Vardarajan Suneeta and 
Dr Eric Woolgar for discussions.  Claire Bordenave and Christophe Renard for  help with computers.  
I acknowledge funding from Peterhouse and from the Killam Foundation at various times during which 
I produced this work.  I thank the organizers, participants and staff at Black Holes V in Banff  
for a pleasant few days during which some of this work was done.



\begin{thebibliography}{99}
\bibitem{Albouy}              J. Barbour suggested to me that the 
                              $N = 4$ $d = 1$ scale invariant problem would be usefully nontrivial, 
                              following earlier discussions of his with A. Albouy.
\end{thebibliography}
\end{document}